\documentclass[fleqn,usenatbib]{mnras}

\usepackage[T1]{fontenc}
\usepackage{ae,aecompl}

\usepackage{graphicx}	
\usepackage{amsmath}	
\usepackage{amssymb}	

\title[Gaseous haloes and the onset of galaxy formation]{The detailed 
structure and the onset of galaxy formation in low-mass gaseous dark matter haloes}

\author[A. Benitez-Llambay et al.]{
Alejandro Benitez-Llambay$^{1}$\thanks{E-mail: alejandro.b.llambay@durham.ac.uk (ABL)}
and Carlos Frenk$^{1}$
\\
$^{1}$Institute for Computational Cosmology, Department of Physics, 
Durham University, South Road, Durham, DH1 3LE, UK\\
}

\date{Accepted XXX. Received YYY; in original form ZZZ}

\pubyear{2020}

\begin{document}
\label{firstpage}
\pagerange{\pageref{firstpage}--\pageref{lastpage}}
\maketitle

\begin{abstract}
We present a model for the formation of the first galaxies before
  and after the reionization of hydrogen in the early universe. In
  this model, galaxy formation can only take place in dark matter
  haloes whose mass exceeds a redshift-dependent critical value,
  which, before reionization, is equal (in the simplest case) to the
  mass at which atomic hydrogen cooling becomes effective and, after
  reionization, is equal to the mass 
  above which gas cannot remain in hydrostatic equilibrium. We define
  the {\em Halo Occupation Fraction} (HOF) as the fraction of haloes
  that host a luminous galaxy as a function of halo mass. The HOF is
  established by the interplay between the evolution of the critical
  mass and the assembly history of haloes and depends on three
  factors: the minimum halo mass for galaxy formation before
  reionization, the redshift of reionization, and the intensity of the
  (evolving) external photoheating rate. Our fiducial model predicts a
  cutoff in the galaxy mass function at a present-day halo mass,
  $M_{200} \sim 3\times 10^{8} M_{\odot}$; 100\% occupation at
  $M_{200} > 5\times 10^9 M_{\odot}$; and a population of starless
  gaseous haloes of present-day mass in the range $10^{6} \lesssim
  M_{200} / M_{\odot}\lesssim 5\times 10^{9}$, in which the gas is in
  thermal equilibrium with the ultraviolet background radiation and in
  hydrostatic equilibrium in the gravitational potential of the
  halo. The transition between ${\rm HOF} = 0$ and ${\rm HOF=1}$
  reflects the stochastic nature of halo mass growth. We explore how
  these characteristic masses vary with model assumptions and
  parameter values. The results of our model are in excellent
  agreement with cosmological hydrodynamic simulations of galaxy
  formation. 
\end{abstract}



\section{Introduction}
\label{Sec:Introduction}
The ``$\Lambda$-cold dark matter'' ($\Lambda$CDM) cosmological model
makes specific and robust predictions for the growth, structure and
abundance of dark matter haloes, the sites where galaxies form. These
are: {\em (i)} dark matter haloes grow hierarchically: small haloes
form first and larger haloes form subsequently by mergers of smaller
haloes and accretion of smooth dark matter; {\em (ii)} the internal
structure of dark matter haloes is self-similar over $\sim 20$ decades
in mass~\citep{Navarro1996, Navarro1997, Wang2019}; {\em (iii)} the
halo mass function rises steeply towards low masses
\citep[approximately as $M^{-1}$; e.g.,][and references
therein]{Press1974, Bond1991, Jenkins2001}, implying that a large
fraction of low-mass haloes must remain ``dark'' at redshift $z=0$ if
$\Lambda$CDM is to be reconciled with the relatively flat faint end of
the observed galaxy stellar mass function~\citep{Klypin1999, Bullock2000,
  Benson2002b,Somerville2002}.

In $\Lambda$CDM a ``cutoff'' in the mass of dark matter haloes that
can host galaxies is expected on general grounds. After the hydrogen
emerging from the Big Bang recombines, it can, in principle, cool and
condense into the small dark matter haloes present at that time but
only in those whose potential well is deep enough to overcome the
kinetic and thermal energy of the gas. Once the hydrogen is reionized
at redshift, $z_{\rm re}$, the hydrogen-helium plasma is heated to
$T_{\rm b} \sim 2 \times 10^{4} \ K$, preventing it from accreting
onto dark matter haloes of virial temperature,
$T_{200} \lesssim T_{\rm b}$~\citep[][]{Efstathiou1992,
  Babul1992,Quinn1996, Thoul1996, Barkana1999}, corresponding to a
mass, $M_{200} \lesssim 10^{10} \ M_{\odot}$\footnote{We define virial
  quantities as those pertaining to the sphere within which the mean
  mass density equals 200 times the critical density of the universe
  and label them with the subscript 200.}. Thus, we expect most haloes
today of mass below a characteristic critical mass, $M_{\rm cr}^0$, to
remain dark. This conclusion agrees qualitatively with results from
cosmological hydrodynamical
simulations~\citep[][]{Hoeft2006,Okamoto2009, Benitez-Llambay2015,
  Sawala2016, Fitts2017}, although the exact value of the critical
halo mass and its dependence on modelling details, are uncertain. A
quantity often discussed in this context is the ``filtering mass''
defined as the mass of those haloes that retain half of their baryons
after cosmic reionization; it can be calculated using either linear
theory or numerical simulations~\cite[e.g.][]{Gnedin2000, Benson2002a,
  Okamoto2008, Hoeft2006}. The exact connection of this quantity to
galaxy formation is, however, unclear.

\begin{figure}
    \includegraphics{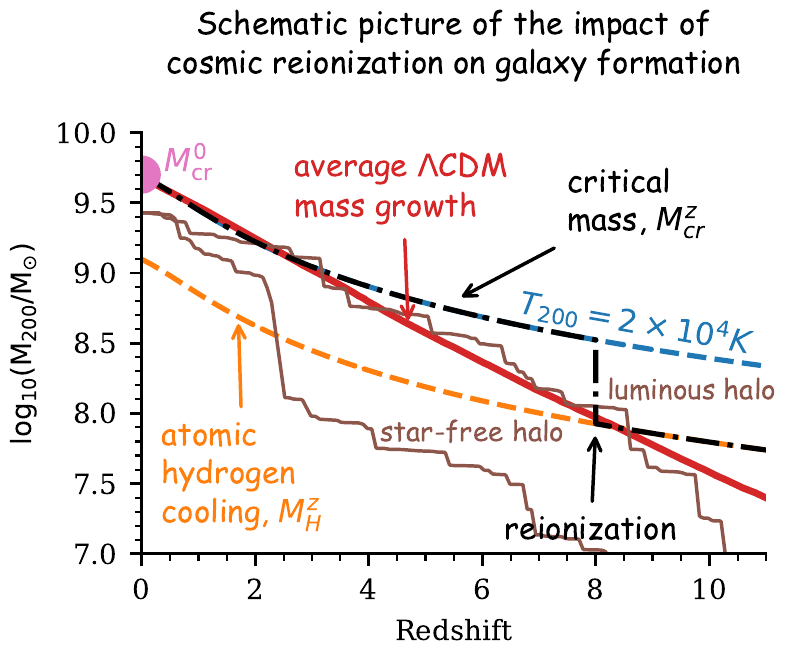}
    \caption[]{Redshift evolution of the critical halo mass above
      which atomic hydrogen cooling becomes efficient (orange
      dashed line), the critical mass corresponding to a fixed virial
      temperature, $T_{200} = T_{\rm b} = 2 \times 10^4 \ K$ (blue
      dashed line), and the mean mass assembly history of a
      $\Lambda$CDM halo of present-day mass,
      $M_{200} \sim 5 \times 10^{9} \ M_{\odot} h^{-1}$ (red solid
      line). The effective critical mass for gas to collapse is
      shown by the dot-dashed black line. This is equal to $M_{H}^z$
      (equation ~\ref{Eq:atomic_hydrogen}) prior to the redshift of
      reionization, $z_{\rm re}$, and to
      $M_{\rm cr}^z$ (equation~\ref{Eq:critical_mass_intro}) after
      $z_{\rm re}$. For this particular example, the value of
      $M_{\rm cr}^0$ constrains both, the redshift of reionization and
      the temperature of the intergalactic medium. The brown thin solid lines show two particular mass assembly histories of dark matter haloes of present-day mass, $M_{200} \sim 3 \times 10^{9} M_{\odot}$, i.e., of mass, $M_{200} < M_{\rm cr}^0$. One halo never exceeds $M_{\rm cr}^z$ and is expected to remain ``dark'' at $z=0$. The other was more massive than $M_{\rm cr}^{z}$ prior to reionization and is expected to host a luminous galaxy at $z=0$.}
    \label{Fig:schematic_reionization}
\end{figure}

To illustrate these ideas, let us assume that prior to reionization
galaxy formation can only take place in haloes in which atomic
hydrogen can cool, that is haloes of virial temperature, 
$T_{200} \gtrsim 7000 \ K$. The corresponding critical mass
is approximately,
\begin{equation}
    M_H^z \sim (4 \times 10^{7} \ M_{\odot} ) \left ( \frac{1+z}{11} \right)^{-3/2}. 
\label{Eq:atomic_hydrogen}
\end{equation}
In Fig.~\ref{Fig:schematic_reionization} we follow the evolution of a
halo of present-day mass,
$M_{\rm cr}^0 = 5 \times 10^{9} \ M_{\odot}$, which, for illustration
purposes, we take to be the critical mass above which {\it all} dark
matter haloes host a luminous galaxy at $z=0$. The red solid line shows the
average mass growth (inferred from the mean mass accretion history) of
a halo of that mass in $\Lambda$CDM.  The halo mass required for
galaxy formation to proceed is shown by the black dashed line. At
$z=z_{\rm re}$ this jumps from $M_{H}^{z}$ to $M_{\rm cr}^z$.  The blue
dashed line in Fig.~\ref{Fig:schematic_reionization} shows the
evolution of $M_{\rm cr}^z$, assumed to be the mass of a halo of virial temperature $T_{\rm b} = 2\times 10^4 \ K$.

{\it All} haloes of $M_{200} \ge M_{\rm cr}^0$ will host a luminous
galaxy at $z=0$. {\it Some} haloes of $M_{200} \leq M_{\rm cr}^0$ will
be ``dark'' but others will also host a luminous galaxy, depending on
their previous history.  The thin brown lines in
Fig.~\ref{Fig:schematic_reionization} illustrate two different mass
accretions histories that lead to the same halo mass at $z=0$,
$M_{200} < M_{\rm cr}^0$.  One of them never crosses the critical mass
required for gas to collapse, whereas the other, although below the
critical mass at $z=0$, exceeded $M_{H}^{z}$ before cosmic
reionization. Of the two, only the later is expected to host a
luminous galaxy today. This example illustrates the origin of the
``stochastic'' nature of galaxy formation in dark matter haloes of
mass close to the critical value~\citep[e.g.][and references
therein]{Hoeft2006, Benitez-Llambay2015, Sawala2016, Fitts2017}.

The present-day value of the critical mass, $M_{\rm cr}^0$, that
separates haloes that were able to form a galaxy from those that were
not is a direct probe of the epoch of reionization and constrains the
two parameters, $z_{\rm re}$ and $T_{\rm b}$.  The redshift of
reionization can be estimated by finding the time when the average
mass of haloes of $M_{200} = M_{\rm cr}^0$ first exceeded $M_{H}^{z}$
(dashed orange line).  The constraint on $T_{\rm b}$ follows from the
value of $M_{\rm cr}^0$ since, 
\begin{equation}
    M_{\rm cr}^z \sim \left ( 10^{10} \ M_{\odot} \right ) \left ( \displaystyle\frac{T_{\rm b}}{3.2 \times 10^{4} \ K} \right )^{3/2}  \left ( 1 + z \right )^{-3/2}.
\label{Eq:critical_mass_intro}
\end{equation}

In this paper we develop a theoretical framework to understand the
onset of galaxy formation and the impact of cosmic reionization on
galaxy formation and address the follow questions: Is there an actual
minimum halo mass at $z=0$ below which galaxies cannot form? If so,
how does this depend on the characteristic scales of cosmic
reionization?  Is the simple picture outlined above consistent with
full hydrodynamical simulations of galaxy formation? Is our model
quantitatively robust?  How sensitive are the results to the
underlying assumptions about galaxy formation?

In Section~\ref{Sec:Model} we introduce our model to calculate how reionization
affects galaxy formation in low-mass haloes. We describe the numerical
method and our simulations in Section~\ref{Sec:Simulations}. We
perform a comparison between our model and high-resolution
cosmological hydrodynamics simulations in
Section~\ref{Sec:Results}. We discuss our results in
Section~\ref{Sec:Discussion} and conclude in
Section~\ref{Sec:Conclusions}.

\section{MODEL}
\label{Sec:Model}

We will discuss later the conditions required for gas to collapse in a
halo and make a galaxy.  Here, we introduce a model to investigate the
effects of reionization and of the presence of an external ultraviolet
background (UVB) on the structure of gaseous haloes. The model applies
{\it strictly} to dark matter haloes whose gaseous component {\it has
  not yet collapsed} to form a galaxy by $z_{\rm re}$.

\begin{figure}
    \includegraphics{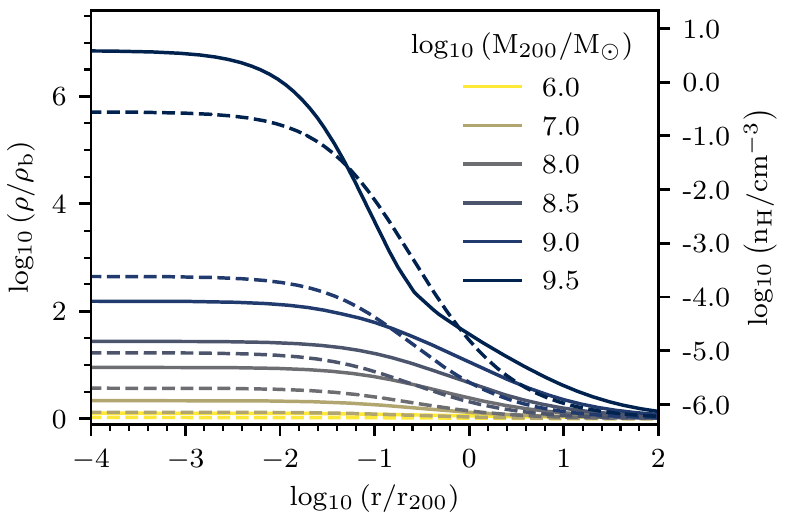}    
    \caption[]{Gas density profile of gaseous haloes in hydrostatic equilibrium with the dark matter potential at redshift $z=0$, for haloes of different mass, as indicated in the legend. The dashed lines show the {\it isothermal} case, for which the temperature of the gas is independent of density and equal to $T = 2\times 10^{4}$. The solid lines correspond to the density-temperature relation, $T(\rho)$, consistent with the gas being in thermal equilibrium with the external~\cite{Haardt2001} UVB radiation field at $z=0$ (see also Fig.~\ref{Fig:eos}).}
    \label{Fig:density_profiles}
\end{figure}

As discussed in the Introduction, cosmic reionization drastically
reduces the gas content of dark matter haloes of virial temperature
$T_{200} \lesssim T_{\rm b} \sim 2 \times 10^{4} \ K$, or
equivalently, mass, $M_{200} \lesssim M_{\rm cr}^z$
(equation~\ref{Eq:critical_mass_intro}). This is the mass scale below
which the halo free-fall time, $t_{\rm ff}$, becomes longer than the sound
crossing time of the gas in the halo, $t_{\rm sc}$\footnote{The condition,
  $t_{\rm ff} < t_{\rm sc}$, yields the well-known criterion for dark matter
  haloes to be affected by cosmic reionization, $V_{200} < c_{s}$ ,
  where $V_{200}$ is the halo circular velocity and
  $c_{s} \sim (10 {\rm \ km/s}) (T_{\rm b}/10^4 K)^{1/2}$ is the
  sound speed of the gas.}. At $z=0$ these haloes have mass,
$M_{200} \lesssim M_{\rm cr}^0 \sim 10^{10} \ M_{\odot}$, but at
$z=10$, the critical mass is
$M_{c}^{10} \sim 5\times 10^{8} \ M_{\odot}$. Any gas that settles in
haloes of virial temperature $T_{200} \lesssim T_{\rm b}$ is
expected to be in thermal (rather than virial) equilibrium with the
external ionizing radiation at a temperature $T\sim T_{\rm b}$.

The distribution of gas that collapses into these low-mass haloes
after reionization can be fully determined assuming that: {\it (i)} the gas
reaches hydrostatic equilibrium within the (spherically symmetric)
dark matter halo, and {\it (ii)} the contribution of the gas to the
gravitational potential is negligible everywhere~\citep[see
e.g.][]{Ikeuchi1986, Rees1986}. Under these assumptions, the
hydrostatic equilibrium equation for the gas reads:
\begin{equation}
    \label{eq:hydrostatic_equilibrium}
    \left ( \displaystyle\frac{\tilde T}{\tilde \rho} + \displaystyle\frac{d\tilde T}{d\tilde \rho} \right ) \displaystyle\frac{d\tilde \rho}{d\tilde r} = -2 \displaystyle\frac{\tilde M(\tilde r)}{\tilde r^2},
\end{equation}
\noindent where $\tilde \rho = \rho / \bar \rho$, $\tilde M = M /
M_{200}$, $\tilde r = r/r_{200}$, $\tilde T = T/T_{200}$, $\bar \rho =
\rho_{c} \Omega_b $ are dimensionless variables; for the virial
temperature, $T_{200}$, we adopt the definition $T_{200} = \mu m_{\rm
  p}/ 2 k_{\rm B} V_{200}^2$, with $\mu$, $m_{\rm p}$ and $k_{\rm B}$
being the mean molecular weight, the proton mass and the Boltzmann
constant, respectively; $V_{200}^2 = G M_{200}/r_{200}$ is the halo
circular velocity at $r_{200}$. We have also made the assumption that
the gas temperature is a function of density, $\tilde T = \tilde
T(\tilde \rho)$.  

\begin{figure}
    \centering
    \includegraphics{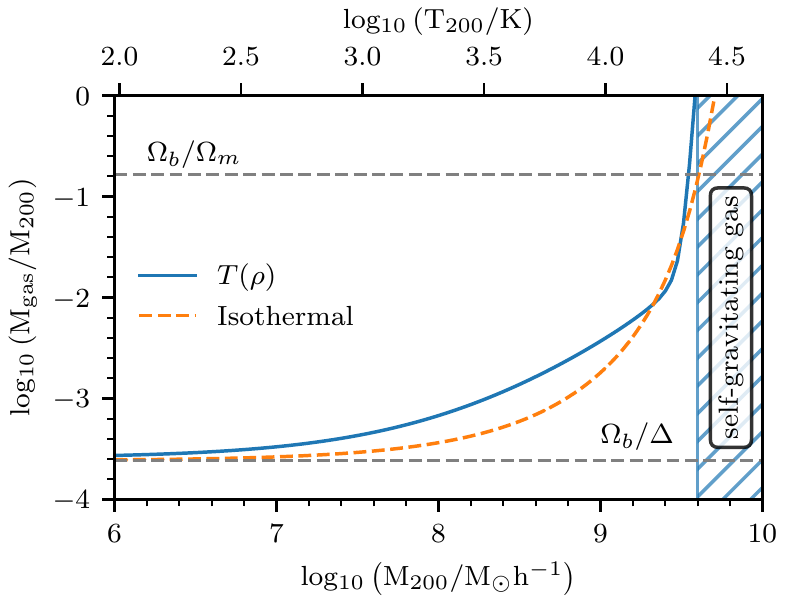}
    \caption[]{Gas mass of starless gaseous haloes, 
    in units of halo mass, as a
      function of halo mass, for the {\it isothermal} case ($T_{\rm b} = 2\times 10^{4} K$; orange
      dashed line) and the general model ($T(\rho)$; solid
      blue line). Gas in haloes more massive than, 
      $M_{\rm cr}^0 \sim 5 \times 10^9 \ M_{\odot} h^{-1}$, at $z=0$
      is expected to become self-gravitating and undergo gravitational collapse. The gas fraction for low-mass haloes approaches the lower bound, $\Omega_{b}/\Delta$, which is the enclosed gas mass inside
      $r_{200}$ of a halo embedded in an unperturbed background. The scale on the top indicates the virial temperature, $T_{200}$.}
    \label{Fig:structural}
\end{figure}

\subsection{Isothermal case} 

For the simplest case in which the temperature of the gas is
independent of density, (i.e., for an {\it isothermal} gas), the
solution of this equation yields:
\begin{equation}
    \tilde\rho(\tilde r) = \exp \left\{ 2 \left ( \displaystyle\frac{T_{200}}{T_{\rm b}} \right ) \displaystyle\int_{\tilde r}^{\infty} \displaystyle\frac{ \tilde M(r')}{r'^2} \ dr'\right\},
\end{equation}
where we assume that $\tilde \rho = 1$ sufficiently far from the halo
($\tilde r\rightarrow\infty$), i.e., the outer pressure of the halo is
equal to that of the intergalactic medium. The enclosed gas mass,  
\begin{equation}
    \tilde M_{\rm gas}(\tilde r) = \left ( \displaystyle\frac{3 \Omega_{b}}{\Delta} \right )  \displaystyle\int_{0}^{\tilde r} \tilde \rho(\tilde r') \tilde r'^2 \ d\tilde r',
\label{Eq:gas_mass}
\end{equation}
in which $\Delta = 200$, must be calculated numerically.

For a dark matter halo described by a Navarro-Frenk-White (NFW)
density profile~\citep{Navarro1996, Navarro1997}, the gas density
profile is:
\begin{equation}
\label{Eq:gas_density_profile_isothermal}
    \tilde\rho(\tilde r) = \exp \left\{ 2 \left ( \displaystyle\frac{ T_{200}}{T_{\rm b}} \right ) \displaystyle\frac{1}{f_c} \displaystyle\frac{\ln(1+c\tilde r)}{\tilde r } \right\},
\end{equation}
in which $f_c = \ln(1+c)-c/(1+c)$, $c=r_{200}/r_s$ is the
concentration of the halo, and $r_s$ is the characteristic radius of
the dark matter profile.

\begin{figure}
    \includegraphics{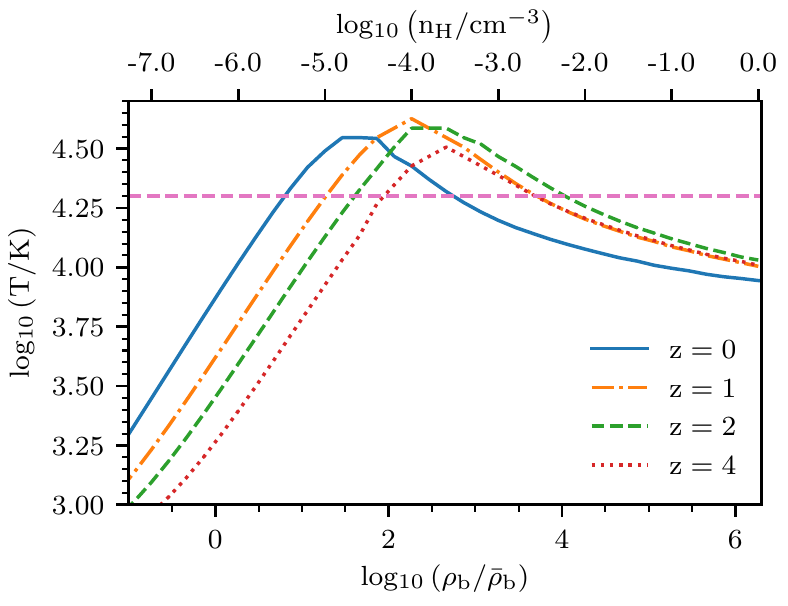}
    \caption[]{Temperature-density relation for the gas in {\it
        starless} dark matter haloes of virial temperature,
      $T_{200} \lesssim 10^4 \ K$. The lines show the $T(\rho)$
      relation expected for gas in thermal equilibrium with the
      redshift-dependent~\cite{Haardt2001} UVB radiation field (see
      Section~\ref{Sec:Model}). The horizontal dashed line shows the
      temperature assumed for the {\it isothermal} case
      ($T_{\rm b} = 2 \times 10^{4} K$), in which the entire gas
      content of the universe is at a single constant temperature,
      independently of density.}
    \label{Fig:eos}
\end{figure}

The dashed lines in Fig.~\ref{Fig:density_profiles} show the gas density
profile (equation~\ref{Eq:gas_density_profile_isothermal}) for
different halo masses, assuming a constant concentration, $c=10$,
independently on halo mass\footnote{Note that the concentration of low-mass
  $\Lambda$CDM haloes is only weakly dependent on halo
  mass~\citep[see, e.g.,][]{Ludlow2014, Wang2019}, so this is a sensible
  assumption for our purposes.}. The slope of the gas density profile
becomes increasingly shallow towards the centre, and converges to a
well-defined value, $\rho_{0}$, at $r=0$. The finite maximum central
gas density is a consequence of the cuspy nature of
$\Lambda$CDM haloes, which gives rise to a maximum central
gravitational acceleration. In addition, the gas density profile
settles at the mean density of the universe ($\tilde \rho = 1$)
further out, by construction.

The total gas mass within the virial radius, as a function of halo
mass (equation~\ref{Eq:gas_mass}), is shown by the dashed orange line
in Fig.~\ref{Fig:structural}. The gas mass grows
exponentially with halo mass and asymptotically approaches the value
$M_{200} \Omega_{b} / \Delta$ for low-mass haloes, as
expected\footnote{The enclosed gas mass within the virial radius must
  scale linearly with halo mass for a flat gas density profile.}. For
more massive haloes, the gas mass becomes a steep function of halo
mass and the amount of gas needed to balance the gravitational
acceleration of the halo eventually exceeds both, the baryon fraction
of the universe (horizontal dashed line) and the dark matter content of the halo,
after which equation (\ref{Eq:gas_mass}) is no longer valid. Gaseous haloes
more massive than $M_{200} \sim 5\times 10^{9} \ M_{\odot} h^{-1}$
will thus become {\it self-gravitating} and continue their
gravitational collapse to form a central luminous galaxy\footnote{Note 
that this is particularly true for dark matter haloes with virial temperature 
$10^{4} \lesssim T_{200} \lesssim 10^{5} \ K$, for which the cooling time is much shorter than the Hubble time, so that the gas in the halo is effectively at $T \sim 10^{4} \ K$.}.

Self-gravitating gaseous haloes develop inner densities that exceed
$10^5$ times the mean baryon density of the universe at $z=0$, or
$\rho/m_{p} = n_{\rm H} \sim 1.0 \rm \ cm^{-3}$ (see scale on the right in Fig.~\ref{Fig:density_profiles}). 
This is encouraging, as this estimate is comparable to
the minimum density above which gas is expected to become {\it
  self-shielding, molecular, and self-gravitating}, a regime in which
our model is not designed to work~\cite[see, e.g.,][]{Sykes2019}. This number also agrees well with
the neutral hydrogen column density, $N_{\rm HI}$, above which damped
Ly$\alpha$ (DLA) systems become rare~\citep[e.g.][]{Schaye2001,
  Krumholz2009, Jorgenson2014}. Indeed, for a halo of mass
$M_{200} = 5\times 10^{9} \ M_{\odot}$, assuming that {\it all}
hydrogen inside the central core is neutral, we obtain a central HI column density $\rm N_{\rm HI} \sim 10^{22} \rm \ cm^{2}$.

\subsection{Thermal equilibrium with the UVB} 

More generally, the interplay between gas cooling
and photoheating from the external UVB links the gas temperature to
its density~\citep[see e.g.,][]{Theuns1998, Barkana1999} and so
equation (\ref{eq:hydrostatic_equilibrium}) must be solved taking this into
account. 

The processes that determine the {\it density-temperature}
relation, $T(\rho)$, of gas in dark matter haloes of virial
temperature, $T_{200} \lesssim 10^{4} K$, depend on the density of the
gas. For high-density gas,
$n_{\rm H} \gtrsim 5\times 10^{-4} \ \rm cm^{-3}$, the cooling and heating
timescales are both shorter than the Hubble time and $T(\rho$) is
determined by the balance between cooling and photoheating. For
low-density gas, the cooling timescale is longer than the Hubble time
and $T(\rho)$ is determined exclusively by the photoheating timescale.

\begin{figure}
    \includegraphics{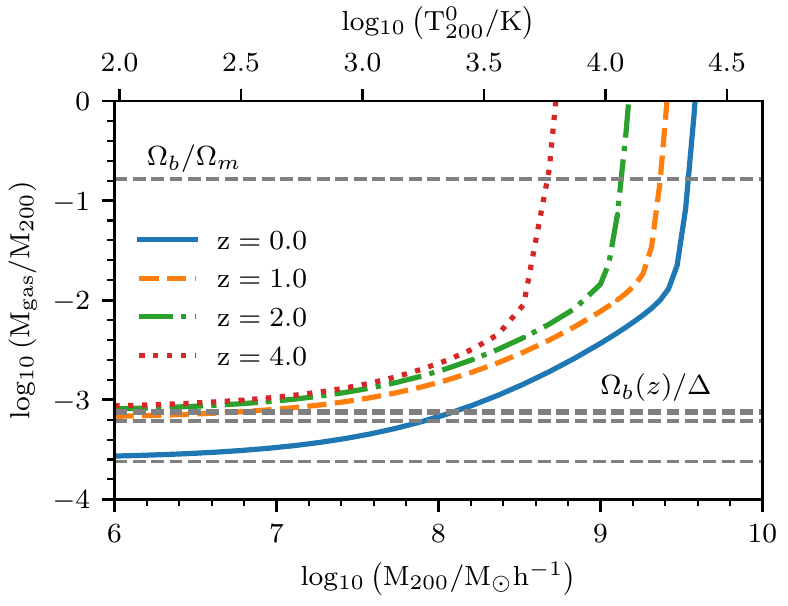}
    \caption[]{Gas mass, in units of the halo mass, as a function of
      halo mass, for {\it starless} gaseous haloes at different
      redshifts. These are derived from the model of
      Section~\ref{Sec:Model}, which assumes that the cosmic gas follows the redshift- and density-dependent $T(\rho,z)$
      relation shown in Fig.~\ref{Fig:eos}. Note that the gas mass
      converges to different values for low-mass haloes as a result of
      the cosmic expansion.}
    \label{Fig:gas_mass_redshift}
\end{figure}

Fig.~\ref{Fig:eos} shows $T(\rho,z)$ calculated for gas of primordial
composition, using the publicly available cooling tables compiled
by~\cite{Wiersma2009}\footnote{Available in
  http://www.strw.leidenuniv.nl/WSS08/}, which assume an optically
thin gas in ionization equilibrium illuminated by
the~\cite{Haardt2001} UVB radiation field. (For completeness, we list
the values of the relation in Table~\ref{Tab:eos} in
Appendix~\ref{App:eos}).  For comparison, the horizontal dashed
line in the figure shows the value of
$T_{\rm b}= 2 \times 10^{4} \ K$ previously assumed for the {\it
  isothermal} case. Although for gas above the mean cosmic density the
temperature is not far from $T_{\rm b}$, there are clear differences.

The $z=0$ {\it density-temperature} relation results in the
gas density profiles shown by the solid lines in
Fig.~\ref{Fig:density_profiles}, and in the present-day gas mass shown by the solid blue line in
Fig.~\ref{Fig:structural}. The results of the simple {\it isothermal}
case follow quite closely those of this model for which the
temperature is not constant. Note, however, that the background
temperature assumed for the {\it isothermal} case, $T_{\rm b}$, is a
free parameter. Had we chosen a different value, the results would
have been different. The $T(\rho)$ relation in Fig.~\ref{Fig:eos} {\it
  removes} any freedom in the modelling.  We will focus on the results
for this general case, in which $T\equiv T(\rho)$, from now on.

Our model exhibits a number of interesting properties: 
\begin{itemize}

\item{There is a wide range in halo mass at $z=0$
    ($10^{6} \lesssim M_{200} / M_{\odot} h^{-1} \lesssim 5 \times
    10^9$) for which gas neither leaves the halo completely nor
    collapses further. We previously identified these
    systems and named them REionization Limited HI Clouds
    (RELHICs)~\citep[][]{Benitez-Llambay2017}. Systems of this kind
    were first explored by~\cite{Rees1986} and~\cite{Ikeuchi1986} in the context of the ``minihalo model of the Lyman-$\alpha$ forest''.}

\item{Dark matter haloes less massive than $M_{200} \sim 10^{6} \
    M_{\odot} h^{-1} $ at $z=0$ are expected to be devoid of gas; 
    their potential wells are not deep enough to balance the pressure
    of the gas.} 
  
\item{Gas in dark matter haloes more massive than
    $M_{200} \sim 5 \times 10^{9} \ M_{\odot} h^{-1}$ at $z =0$ cannot
    provide enough pressure support to prevent further gravitational
    collapse; these haloes are expected to collapse, fragment, and
    form stars.}
    
\end{itemize}

\subsection{Redshift evolution}

The redshift evolution of the model is determined by the time
variation of the external pressure of the intergalactic medium, as the
model assumes that the external pressure is equal to that of the mean
intergalactic medium: $P(z) \propto \rho_{b}(z)T_{\rm b}(z)$. Thus,
the external pressure for the gas in the halo must be calculated
taking into account the redshift evolution of the~\cite{Haardt2001}
photoheating rate (see Fig.~\ref{Fig:eos}) and the cosmic expansion.

As an example, Fig.~\ref{Fig:gas_mass_redshift} shows the gas mass (in
units of the halo mass), as a function of halo mass, at four different
redshifts, $z=(0,1,2,4)$.  It is clear that, at fixed halo mass, the
total gas mass of these low-mass haloes evolves with time. At high
redshift, the universe is denser and so the external pressure on the
gaseous haloes is greater. As a result, the critical mass above which
gas is expected to lose pressure support shifts to lower halo
masses. This invites us to define the {\it redshift-dependent}
critical mass for gas to undergo gravitational collapse,
$M_{\rm cr}^z$, as the halo mass above which the fraction of baryons
needed to ensure hydrostatic equilibrium exceeds the universal
fraction, $f_{b} = \Omega_{b}/\Omega_{m}$, indicated in the figure as
a dashed line. Clearly, the exact value of this fraction is not
crucial as long as
$\tilde M_{\rm gas} = M_{\rm gas}/M_{200} \gtrsim 0.1$.

\begin{figure}
    \includegraphics{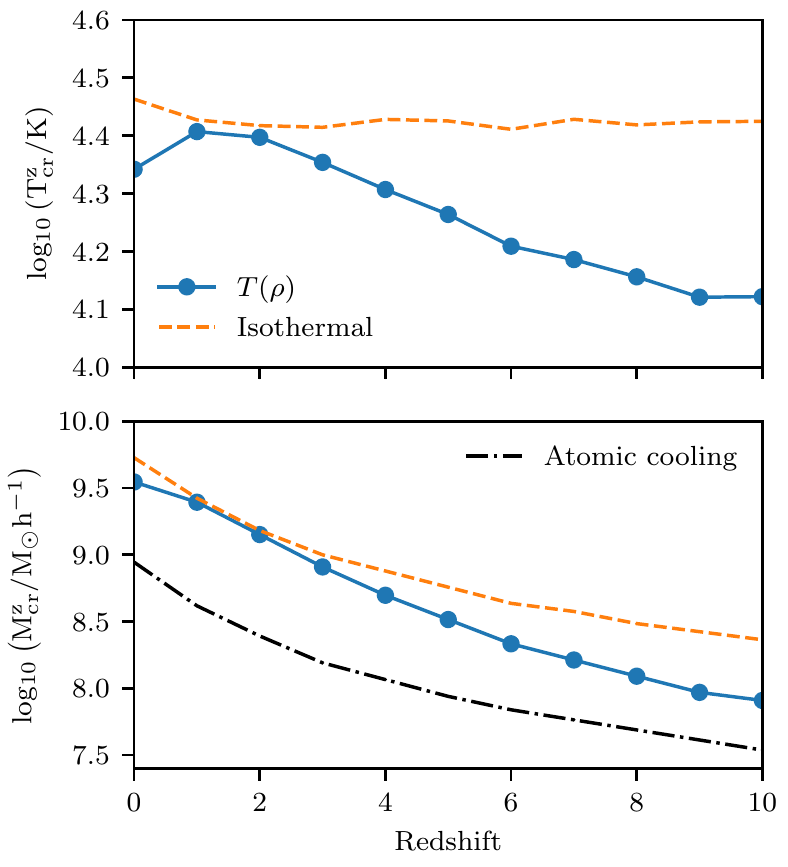}
    \caption[]{Critical virial temperature (top) and critical gas mass
      (bottom), as a function of redshift. The orange dashed line
      shows the result for the {\it isothermal} case, assuming
      $T_{\rm b} = 2 \times 10^4 K$. The solid blue line shows the
      result of the general model of  Section~\ref{Sec:Model}, which
      has no free parameters (except for the intensity of the UVB
      radiation field). The black dot-dashed line in the bottom panel
      shows the critical halo mass above which atomic hydrogen cooling
      becomes efficient (equation~\ref{Eq:atomic_hydrogen}).}
    \label{Fig:critical_mass}
\end{figure}

The blues lines in Fig.~\ref{Fig:critical_mass} show the redshift
evolution of the critical mass that results from our model, expressed
either as a virial temperature (top), or as a gas mass
(bottom)\footnote{For $z>10$, we use the~\cite{Haardt2001} UVB
  background that corresponds to the earlier tabulated redshift,
  $z=10$.}. Note that the critical virial temperature for gas to
become {\it self-gravitating} and collapse is not a constant, contrary
to common assumptions~\citep[e.g.][]{Bullock2000, Okamoto2009,
  Bose2019, Graus2019}. Moreover, the critical temperature exhibits a
maximum at $z\sim 2$, when the~\cite{Haardt2001} photoheating rate
peaks.

For comparison, we also show the critical mass and virial
temperature for the particular {\it isothermal} case in which 
$T_{\rm b} = 2\times 10^{4} K$. For this particular case, the
critical temperature remains constant, as expected. Interestingly,
for this particular value of $T_{\rm b}$, the {\it isothermal}
case resembles our model reasonably well. Note, however, that it can
depart from our model by more than a factor of $3$ at high redshift.
The black dashed line in the bottom panel shows the critical halo
mass above which atomic hydrogen cooling becomes efficient
(equation~\ref{Eq:atomic_hydrogen}), and corresponds 
to the critical halo mass for galaxy formation in the absence of an
external UVB radiation field. 

We conclude this section by highlighting another central feature of our model:
\begin{itemize}
\item {Dark matter haloes less massive than
    $\sim 5 \times 10^{9} \ M_{\odot} h^{-1}$ at $z=0$ that did not exceed either the atomic hydrogen cooling limit prior to cosmic reionization   or the critical mass imposed by the external UVB afterwards will remain ``dark'' at $z=0$; otherwise they will be luminous (see also
    Fig.~\ref{Fig:schematic_reionization}). Galaxy
    formation below the (present-day) mass scale
    $M_{\rm cr}^0 \sim 5 \times 10^{9} M_{\odot} h^{-1}$ is thus largely {\it stochastic} and dependent on the assembly
    history of individual $\Lambda$CDM haloes.}
\end{itemize}

We now turn our attention to a comparison between the predictions of
our models and the results of high-resolution hydrodynamical
cosmological simulations. We introduce the simulation in
Section~\ref{Sec:TheSimulation}, provide details of the identification
of substructures in Section~\ref{Sec:HBT}, and present the detailed
comparison in Section~\ref{Sec:Comparison}.

\begin{figure*}
    \centering
    \includegraphics{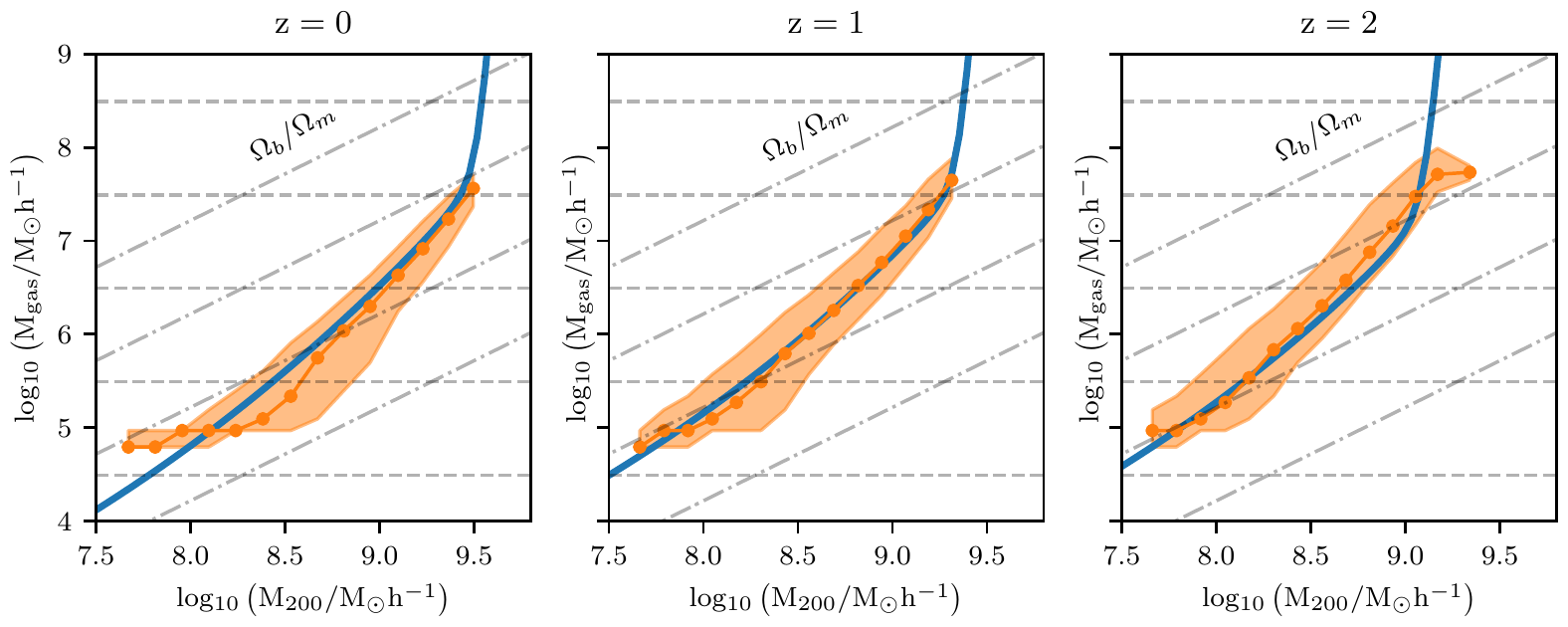}
    \caption[]{Gas mass within the virial radius of simulated {\it starless} gaseous haloes (RELHICs) as a
      function of halo mass at redshifts $z = 0, 1$ and 2. The blue solid
      lines show the gas mass derived from the model of
      Section~\ref{Sec:Model} (see Fig.~\ref{Fig:gas_mass_redshift}),
      which assumes gas in thermal equilibrium with
      the~\cite{Haardt2001} UVB radiation field and in hydrostatic
      equilibrium within its dark matter halo. The running median and the 10-90th percentiles of the simulation data are shown by orange circles and the shaded regions, respectively. Note that our model has no free
      parameters other than the intensity and shape of the UVB
      background. The agreement with the simulation results, and the
      correct redshift-dependence, are remarkable. The oblique dot-dashed grey lines correspond to multiples of the universal baryon fraction, $\alpha\times \Omega_{b}/\Omega_{m}$, with $\alpha=(1,0.1,0.01,0.001)$. The horizontal dashed lines indicate, from top to bottom, the gas mass that corresponds to  $10^{4}, 10^{3}, 10^{2}, 10$, and $1$ gas particles.}
    \label{Fig:gas_mass_sim}
\end{figure*}

\section{Techniques}
\label{Sec:Simulations}

\subsection{Simulation details}
\label{Sec:TheSimulation}
We consider a random realisation of a periodic cosmological cubic
volume of side $20 \rm \ Mpc$, with initial conditions obtained using
the publicly available code {\tt MUSIC}~\citep{Hahn2011}. The
simulation volume is filled with $1024^3$ dark matter particles and
the same number of gas particles, so that the dark matter and gas
particle masses are: $m_{\rm drk} = 1.98 \times 10^5 M_{\odot} h^{-1}$
and $m_{\rm gas} = 3.68 \times 10^{4} M_{\odot} h^{-1}$,
respectively. The Plummer-equivalent gravitational softening, $\epsilon$, adopted in our simulation never exceeds $1 \%$ of the mean interparticle separation. This gives $\epsilon \sim 195 \rm \ pc$ for both the gas and the dark matter particles.  We assume cosmological parameters consistent with early
Planck results~\citep{Plank2014}\footnote{We have chosen these
  parameters so that our simulation is consistent with those of the
  EAGLE collaboration.}.

We performed the simulation with the version of the code {\tt
  P-Gadget3}~\citep[last described in ][]{Springel2005} previously
used for the EAGLE project~\citep{Schaye2015, Crain2015}. We refer the
reader to the original papers for further details about the physics 
modules included in the simulation. We describe here only those relevant
for our analysis.

The simulation includes radiative cooling and a uniform, isotropic, 
redshift-dependent UVB spectrum~\citep{Haardt2001}. The simulation
follows star formation, stellar evolution, supernova feedback and
chemical enrichment. Gas cooling is calculated using
the~\cite{Wiersma2009} cooling tables, which include cooling due to 
several metal species. For our purposes, however, cooling by 
primordial hydrogen and helium are the only relevant processes.

Reionization is modelled by turning on the isotropic
~\cite{Haardt2001} UVB at the redshift of reionization, assumed to be
$z_{\rm re} = 11.5$\footnote{Recent results~\citep{Planck2018}  are consistent with a redshift of reionization, $z_{\rm re} \sim 7.5$. We show in Sec.~\ref{Sec:Results} that our results are not significantly affected by the choice of $z_{\rm re}$ provided $z_{\rm re} \gtrsim 6$.}. In order to ensure that gas is quickly heated to
$\sim 10^4 \ K$ at high redshift, an energy of 2~eV per proton mass is
injected instantaneously at $z=z_{\rm re}$.

Star formation proceeds stochastically according to the
Kennicut-Schmidt law in gas particles whose density is above a
threshold, $n_{\rm H,0}$. In the EAGLE simulations, the density
threshold for star formation is a function of metallicity. We do not
make this assumption here and instead assume a constant value,
$\rm n_{\rm H,0}=1.0 \rm \ cm^{-3}$, motivated by the discussion of
Section~\ref{Sec:Model} (see also Fig.~\ref{Fig:density_profiles}). The results of our simulation should not be strongly dependent on the assumed value of $\rm n_{\rm H,0}$ provided this value is comparable to or larger than $\rm n_{\rm H,0} \sim \rm 1 \ cm^{-3}$, above which gas becomes self-gravitating in low-mass haloes~\cite[e.g.,][]{Benitez-Llambay2019}. Once the gas in the simulation becomes self-gravitating, its maximum density is limited either by resolution or by the adopted threshold.

The EAGLE model imposes a temperature floor,
$T(\rho) = T_{0} \left ( n_{\rm H} / n_{H,0} \right )^{\gamma-1}$ for
high-density gas, $n_{\rm H}>n_{\rm H,0}$, where $\gamma = 4/3$ is the
ratio of specific heats, and $T_0 = 8000 \ K$. We adopt this
temperature floor to prevent the development of extremely high
densities, which might impact the run time of the
simulation. This density-dependent temperature floor affects only
star-forming gas, so it should not affect the comparison between the
{\it starless} haloes of our model and those from the simulation.

\begin{figure*}
    \centering
    \includegraphics{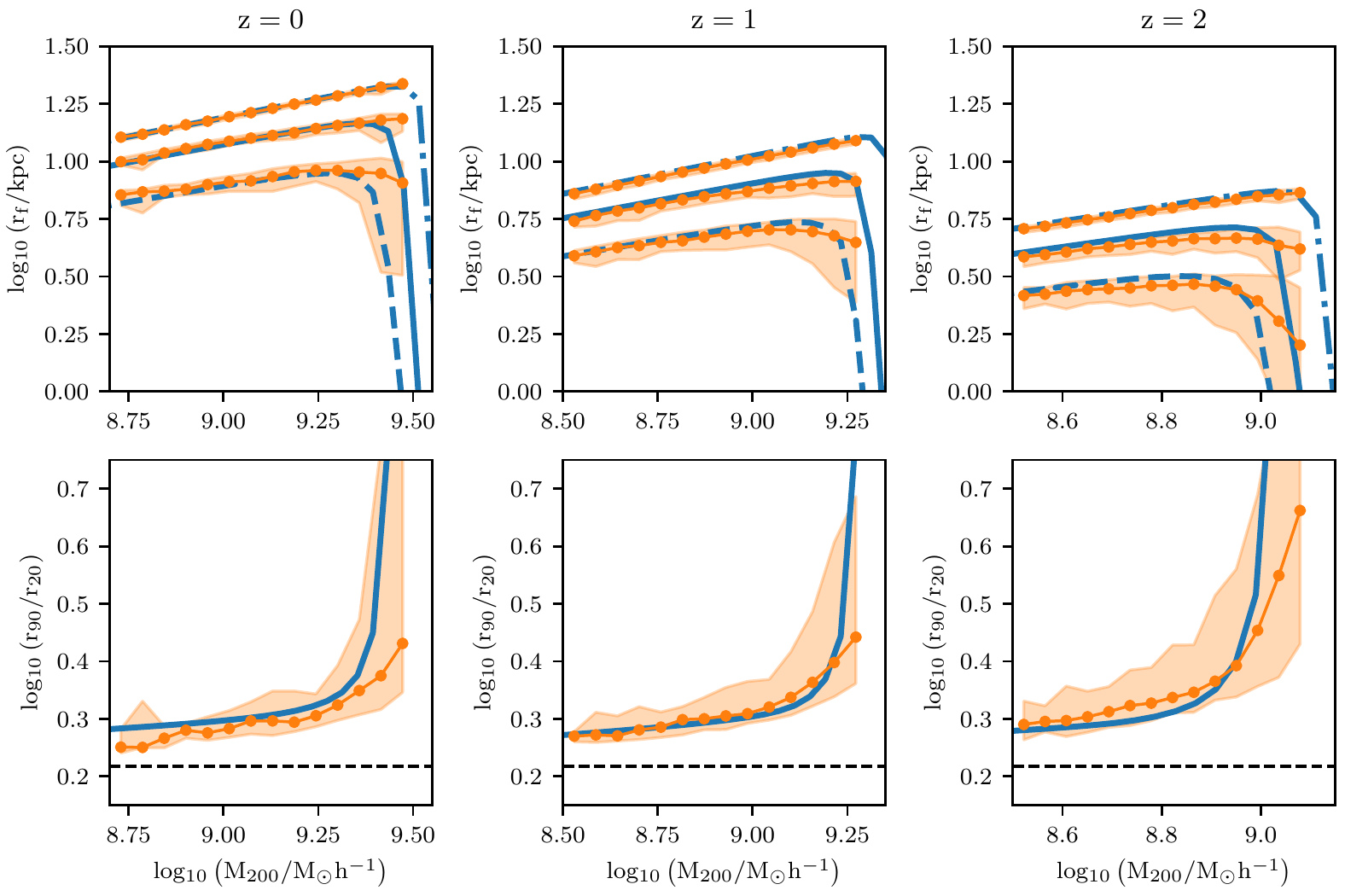}
    \caption[]{{\em Top panels}: the radius, $r_{\rm f}$, encompassing different fractions, $f$, of the total gas mass of {\it starless} gaseous haloes (RELHICs) that contain more than 100 gas particles, as a function of
      halo mass and redshift. From top to bottom the different blue curves show the result of our analytical model of Sec.~\ref{Sec:Model} for $f=(0.9,0.5,0.2)$. The orange circles and shaded regions show the running median and 10-90th percentiles of the distributions, respectively. Different panels correspond to three different
      redshifts as indicated. {\em Bottom panels}: concentration, $r_{90}/r_{20}$, of the gaseous profiles, also as a function of halo mass and redshift. For comparison, the horizontal dashed lines in the bottom panels show the trivial relation expected for a constant gas density profile (i.e., $r_{90}/r_{20} = (9/2)^{1/3}$).
      The agreement between the simulated RELHICs and the analytical model below the critical mass is remarkable.}
    \label{Fig:radii_simulation}
\end{figure*}

\subsection{Identification of haloes in the simulation and mass assembly histories}
\label{Sec:HBT}

Dark matter haloes are identified using the publicly available code
{\tt HBT+}~\citep{Han2018}. In short, {\tt HBT+} uses an input
catalogue of friends-of-friends haloes, constructed using the standard
linking length, $b=0.2$ \citep{Davis1985}, and determines which
particles are gravitationally bound to each halo, discarding the
rest. The unbinding procedure starts at the earliest snapshot ($z=20$
in our case) and the list of gravitationally bound particles in each
halo is then passed to the next snapshot to identify the descendant
halo. When multiple progenitors end up in the same descendant, a main
progenitor is determined according to mass, and the remaining
particles are subject to the unbinding procedure to create satellite
subhaloes. Finally, the bound particles that do not belong to any
satellite are assigned to the main progenitor. These steps are
repeated between consecutive snapshots until the final snapshot of the
simulation is reached.

Applying this algorithm to all the simulation outputs results in a
catalogue of {\it central} and {\it satellite} galaxies at redshift
$z=0$ along with their time evolution. We shall refer to the evolution
of the mass of the main progenitor as the {\it Mass Assembly History}.
High completeness of the $z=0$ halo and subhalo catalogues can be
achieved by sampling the evolution from redshift $z=20$ to $z=0$ with
60 snapshots~\citep{Han2012}. Our simulation has more than twice this
number.

The {\tt HBT+} catalogue provides an extensive set of halo properties,
such as position, velocity, virial mass, gas mass, stellar mass, etc. It
also provides a list of particles (gas, dark matter and stars)
gravitationally bound to each central and satellite halo.  We shall
refer to these particles as the {\it bound} particles.

\subsection{Comparison between our model and our $\Lambda$CDM
  hydrodynamics simulation}
\label{Sec:Comparison}

The properties of RELHICs (REionization Limited HI Clouds;
\citealt{Benitez-Llambay2017}) --{\it starless} haloes whose gas is
in thermal equilibrium with the external UVB radiation field and in
hydrostatic equilibrium within the halo-- provide an important test of
our model. The model (Section~\ref{Sec:Model}) predicts the existence
of these RELHICs in the mass range
$10^{6} \lesssim M_{200} / M_{\odot} h^{-1} \lesssim 5 \times 10^9$ at
$z=0$. In order to test the model we select {\it all} {\it field}
dark matter haloes in our simulation of mass,
$M_{200} \ge 3 \times 10^{7} M_{\odot} h^{-1}$, that contain at least one {\it
  bound} gas particle and {\it no stars}. These haloes are resolved with more than 150 dark matter particles in our simulation.

Fig.~\ref{Fig:gas_mass_sim} shows the {\it bound} gas
mass within the virial radius of the selected haloes, as a function of
halo mass at redshifts $z=0, 1$ and $2$. The running medians and the 10-90th percentiles of the distributions are shown by the orange circles and shaded regions, respectively.

The overall agreement between the gas mass of the simulated RELHICs and the predictions of our model is
remarkable, even in systems that contain fewer than 10 {\it bound} gas particles. Both, the zero point and the slope of the relation
agree. This is not entirely surprising when we consider that the thermodynamic properties of the gas inside {\it starless} systems are largely established by external processes. One of these is the presence of an external gravitational potential sourced by (well-resolved) dark matter haloes; the other is the interplay between photoheating and cooling. The agreement between our model and the simulation demonstrates that the gas particles inside the RELHICs trace the local gas density fluctuations of the intergalactic medium induced by the presence of a dark matter halo.

This result extends those of~\cite{Benitez-Llambay2017}, where
we probed this regime using a limited sample of haloes taken from a
zoom-in simulation of the Local Group at redshift $z=0$ only. Our
simulation validates the assumptions made in Section~\ref{Sec:Model}
in a cosmological context, for a large sample of haloes as a function
of time. Interestingly, the number of gaseous haloes decreases
dramatically as the halo mass approaches the critical mass, where the
model predicts a sharp upturn in gas mass (see
Fig.~\ref{Fig:gas_mass_redshift}). This is because the fraction of dark
matter haloes that host a luminous galaxy increases steadily with halo
mass above $M_{200} > 3 \times 10^{8} M_{\odot} h^{-1}$ and becomes
unity above the critical mass. As a consequence, the number of {\it
  starless} gaseous haloes drops significantly near the critical
mass. This validates our interpretation that the gas in RELHICs is
expected to lose pressure support and collapse once the halo mass
$M_{200} \gtrsim M_{\rm cr}^z$. We will return to this point in
Section~\ref{Sec:Results}.

A further test of the model is provided by the detailed radial
structure of the RELHICs, which we characterise by the radii, $r_{f}$,
encompassing a fraction, $f$, of the total gas mass. For the haloes in the simulation, $r_{f}$ is obtained from the gas particles {\it bound} to the systems while for the model, $r_{f}$ is calculated numerically. In order to measure $r_{f}$ reliably in the simulation we restrict our analysis to systems that contain more than 100 {\it bound} gas particles at the  redshift of interest.

 The top panel of Fig.~\ref{Fig:radii_simulation} shows $r_{f}$, for $f=(0.9, 0.5, 0.2)$, as a function of halo virial mass at the same three
redshifts as before. The orange circles and shaded regions show the running median and 10-90th percentiles for
the simulated haloes. The blue lines show the results of our model. The agreement between the simulated density profiles and the
predictions of the model of Section~\ref{Sec:Model} is very good. This is a non-trivial result that further validates the model and demonstrates that the gaseous structure of RELHICs is well understood. The bottom panel of the same figure shows the concentration of RELHICs, defined as the ratio, $r_{90}/r_{20}$, as a function of halo mass. For comparison, the horizontal dashed line shows the expected concentration if the gas density profile were constant, i.e., $r_{90} = (9/2)^{1/3} r_{20}$.

We conclude that the detailed structure of RELHICs, i.e., post-reionization {\it starless} gaseous haloes in $\Lambda$CDM, can be understood from three simple considerations:
\begin{itemize}
\item{The dark matter profiles of dark matter haloes are well described by the NFW form.}
\item{Gas in RELHICs is in thermal
    equilibrium with the external UVB radiation field.}
\item{The external pressure of RELHICs is equal to that of the intergalactic medium.}
\end{itemize}

\subsection{Synthetic Mass Assembly Histories}

The limited resolution of our simulation precludes following the
formation of dark matter haloes below a minimum mass of a few
$10^{6} \ M_{\odot} h^{-1}$. This imposes a maximum (mass-dependent)
redshift above which our simulation is no longer valid. In order to
overcome this limitation we construct a set of synthetic assembly
histories using the Extended Press-Schechter
formalism~\cite[EPS,][]{Bond1991,Bower1991} and the method proposed
by~\cite{Lacey1993}. Specifically, for 50 $z=0$ mass bins equally
spaced logarithmically in the range
$10^6 \le M_{200} / M_{\odot} h^{-1} \le 10^{12}$, we calculate 500
$\Lambda$CDM assembly histories per mass bin, from redshift $z=36$ to
$z=0$, in 406 equally-spaced redshift bins. We shall refer to these
synthetic assembly histories as ``{\it EPS assembly histories}''.

As we demonstrate in Appendix~\ref{App:MAHs}, our set of EPS assembly
histories agree well with those measured in the simulation down to the
resolution limit of the simulation
($M_{200} \sim 10^{6} M_{\odot} h^{-1}$).

\section{RESULTS}
\label{Sec:Results}

\subsection{Mass assembly histories and the onset of galaxy formation}

Prior to reionization, we will assume that galaxies can only form in
haloes in which atomic hydrogen cooling becomes efficient, as
discussed in the Introduction. (We will test the impact of a different
assumption in Section~\ref{Sec:impact_of_early_gf}.) Once the universe
is reionized, we can follow the fate of the gas in haloes using our
model of Section~\ref{Sec:Model}.  Gas is not able to collapse into every dark matter halo so not all haloes will host a
visible galaxy. We define the {\it Halo Occupation Fraction} (HOF) as
the fraction of dark matter haloes that host a luminous galaxy,
  as a function of redshift and halo mass. Similar definitions have been adopted by other authors to quantify the ability of dark matter haloes to host stars prior to reionization~\citep[e.g.,][]{Xu2016}.
\begin{figure}
    \centering
    \includegraphics{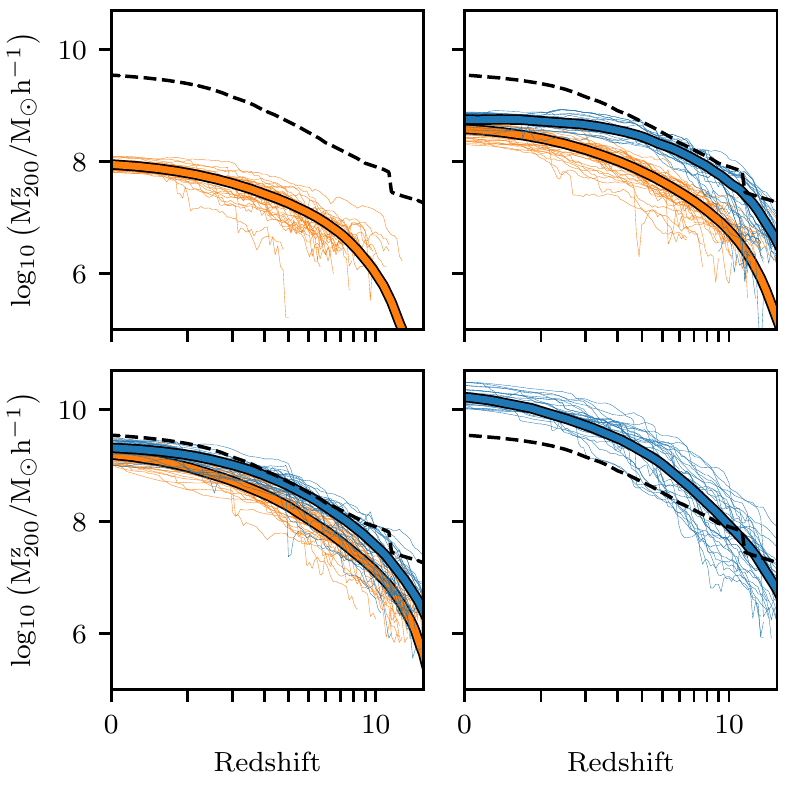}
    \caption[]{Mass assembly histories of halos in our cosmological
      hydrodynamics simulation in four bins of present-day halo mass. The
      thin orange lines correspond to a random sample of {\it field}
      dark matter haloes that do not host a luminous galaxy at
      $z=0$. The thin blue lines correspond to haloes that do contain
      a galaxy, i.e., more than one stellar particle, at their centre. The thick colour lines show the median
      of each set. The black dashed line indicates the critical mass for
      galaxy formation derived from our model 
      (Section~\ref{Sec:Model}). The limit imposed by our model and the
      outcome of the simulation agree remarkably well, suggesting
      that the onset of galaxy formation can be understood in terms of
      our simple model.}
    \label{Fig:simulated_MAHs}
\end{figure}

The onset of galaxy formation is closely related to the assembly
history of $\Lambda$CDM haloes and to the redshift-dependent critical
mass for gas to lose pressure support and collapse.  As we have seen,
after reionization the critical mass is set by the presence of the
UVB radiation field; before reionization further assumptions are
required. 

Fig.~\ref{Fig:simulated_MAHs} shows the assembly history of a random
selection of haloes identified in our $\Lambda$CDM cosmological
hydrodynamics simulation, split into four bins according to their mass
at $z=0$. A variety of different formation histories can give rise to
the same final mass.  The black dot-dashed line shows the critical
mass for galaxy formation derived in Section~\ref{Sec:Model}, assuming
a constant concentration, $c=10$, for all haloes, a redshift of
reionization, $z_{\rm re} = 11.5$ (i.e. consistent with that in our
simulation) and a critical mass for galaxy formation prior to cosmic
reionization described by equation (\ref{Eq:atomic_hydrogen}). Haloes that
do not host a luminous galaxy at $z=0$ are shown by orange lines;
those that have a stellar component (of at least one star particle) at
$z=0$ are shown by blue lines. The medians of each set are shown by
the thick lines of the corresponding colour.

The top left panel shows that the mass of {\it all} dark matter haloes
with present-day mass, $M_{200} \sim 10^{8} \ M_{\odot} h^{-1}$,
always remain below the critical mass required for galaxies to form.
As expected, {\it none} of these haloes hosts a galaxy in our
simulation. More massive haloes may or may not cross the critical mass
for galaxy formation; only those whose mass exceeded, or was
comparable to the critical mass some time in the past will host a
luminous galaxy at $z=0$. Remarkably, we find in the simulation that
indeed the only dark matter haloes that contain a galaxy at their
centre at $z=0$ are those whose mass was above or close to the
critical mass in the past (see, e.g., top-right panel). Moreover, we
find that galaxy formation actually begins soon after the haloes
cross this critical mass (see Appendix~\ref{App:when_galaxies_form}).

As the present-day halo mass increases, so does the fraction of haloes
that cross the critical mass. If the mass today is large enough so
that {\em every} dark matter halo once exceeded the critical mass,
then the fraction of dark matter haloes hosting a luminous galaxy
today becomes unity. This trend was already seen in a number of
studies of the formation of dwarf galaxies in cosmological
hydrodynamical simulations~\cite[e.g.,][and references
therein]{Hoeft2006,Okamoto2008,Okamoto2009, Sawala2016,
  Benitez-Llambay2015, Fitts2017}, but the exact value of the critical
mass for galaxy formation is somewhat uncertain, although it is not
far from the mass corresponding to a virial temperature,
$T_{200} \sim 10^4 \ K$. This uncertainty is removed by our model
which can be used to understand quantitatively the Halo Occupation
Fraction.

The key to understanding the HOF is the mass accretion history. To demonstrate this we
select {\it all} {\it field} dark matter haloes in our simulation 
identified by {\tt HBT+} and compare their growing mass, determined by
their individual assembly histories, to the critical mass,
$M_{\rm cr}^z$. If the mass of a simulated halo exceeds the critical mass
at some redshift, we label this halo as {\it luminous}, that is, we
assume that gas will be able to collapse and form a galaxy at its
centre. Otherwise we label the halo as {\it dark}.  We then calculate
the HOF as a function of halo mass, based on this 
classification only and compare it to the fraction of halos
that actually host a galaxy in the simulation. 

\begin{figure}
    \centering
    \includegraphics{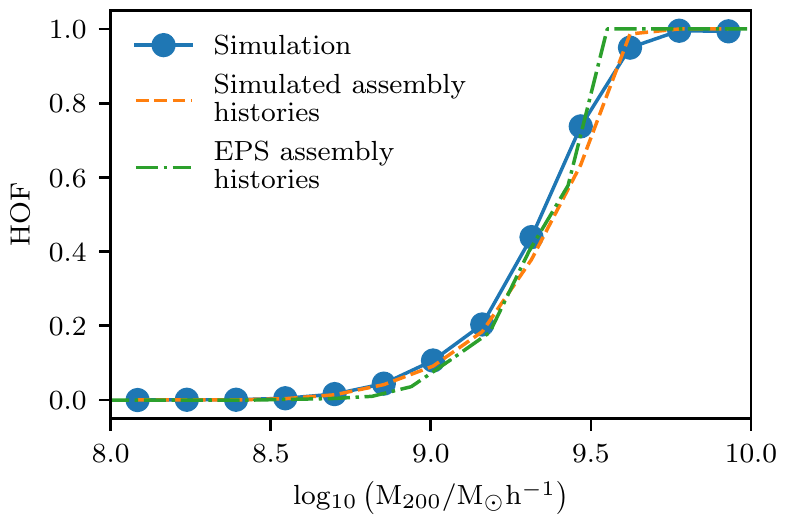}
    \caption[]{Comparison of the $z=0$ halo occupation fraction (HOF)
      measured directly in our cosmological simulation (blue
      circles), derived from the actual halo assembly histories
      measured in the simulation (orange dashed line) and derived
      assuming EPS assembly histories (green dot-dashed
      line). The agreement amongst all the curves demonstrates that
      the onset of galaxy formation in our cosmological hydrodynamical
      simulation is correctly described by the simple model of
      Section~\ref{Sec:Model} coupled to the assembly histories of
      $\Lambda$CDM haloes.}
    \label{Fig:simulated_HOF}
\end{figure}

Fig.~\ref{Fig:simulated_HOF} compares the HOF measured in three ways:
directly in our cosmological simulation at $z = 0$, from our model
applied to the assembly histories inferred in the simulation and from
our model applied to synthetic assembly histories derived from EPS
theory.  The agreement amongst all  three is excellent. This
demonstrates that the HOF is a result of the existence of a critical
halo mass for galaxies to form and the stochastic nature of the growth
of $\Lambda$CDM haloes.  We thus conclude the following (where the
numbers are valid for the \citealt{Haardt2001} UVB assumed in our
simulation):
\begin{itemize}
\item{ At $z=0$ {\it field} dark matter haloes less massive than
    $3\times 10^{8} M_{\odot} h^{-1}$ do not host a luminous galaxy. The gravitational force in these
    haloes is not strong enough to overcome the pressure of the gas
    imparted by the UVB.}
    \item{All dark matter haloes more massive than
        $M_{200} > 5 \times 10^{9} M_{\odot} h^{-1}$ host a galaxy at
        $z=0$. Gas pressure cannot prevent gravitational
        collapse. Once dark matter haloes exceed this mass, they will
        inevitably form a galaxy at their centre.}
    \item{The {\it halo occupation fraction} in the mass range
        $3\times 10^{8} \lesssim M_{200} / M_{\odot} h^{-1} \lesssim 5
        \times 10^{9}$ is determined by the scatter in the {\it
          assembly histories} of $\Lambda$CDM haloes.} 
\end{itemize}

Two main assumptions lie behind these conclusions: {\em (i)} galaxy
formation is largely determined by atomic hydrogen cooling prior to
cosmic reionization; {\em (ii)} cosmic reionization occurred at $z_{\rm re} = 11.5$. We
now discuss how sensitive our conclusions are to variations of these
assumptions. First we consider the impact of varying the redshift of
reionization and then of relaxing our criterion for galaxy formation
in low-mass haloes prior to reionization. To this end we will use our
set of EPS assembly histories.

\subsection{The effect of $z_{\rm re}$ on the $z=0$ Halo Occupation Fraction}

To assess the effect of the redshift of reionization, $z_{\rm re}$, on the
HOF at $z=0$, we use our set of EPS assembly histories and vary the
value of $z_{\rm re}$. This has a direct impact on the value of the
critical mass for galaxy formation. For $z>z_{\rm re}$, we assume that
galaxy formation occurs only in haloes in which atomic hydrogen
cooling is efficient. We model the effect of reionization by assuming
that the cosmic gas is suddenly placed on the {\it density-temperature} relation shown in Fig.~\ref{Fig:eos}, for the
appropriate redshift.

\begin{figure}
    \centering
    \includegraphics{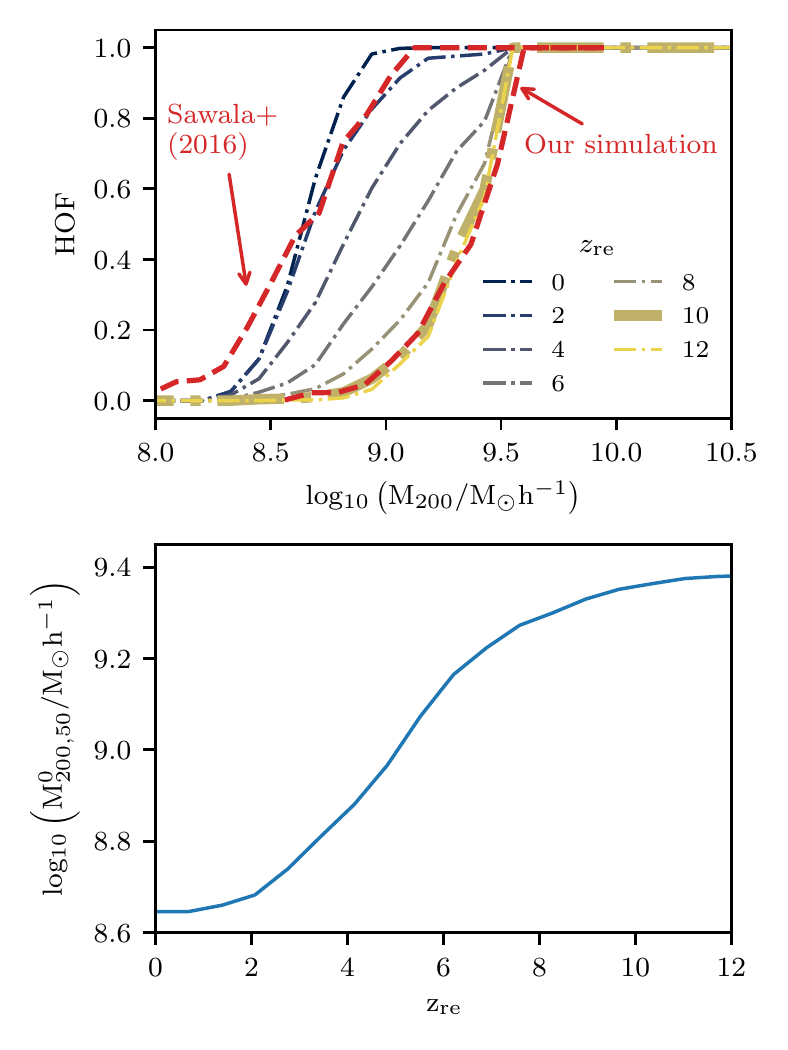}
    \caption[]{{\em Top panel}: the effect of varying the redshift of reionization on the $z=0$
      {\it halo occupation fraction}. If reionization occurred
      early, $z_{\rm re} \gtrsim 10$, the $z=0$ HOF is largely insensitive
      to the exact value of $z_{\rm re}$. On the other hand, if
      $z_{\rm re} \lesssim 10$, then $z_{\rm re}$ has a dramatic effect on the
      $z=0$ HOF. In a universe that never experienced reionization and
      in which galaxies formed through atomic hydrogen cooling, {\it
        all} haloes above $M_{200} \gtrsim 10^{8.9} M_{\odot} h^{-1}$
      would host a luminous galaxy. As a comparison, we show the
      result of a zoom-in simulation of the Local Group
      by~\cite{Sawala2016} that does not include reionization (left red
      dashed line). The agreement between that simulation and the
      predictions using our framework is reassuring. We also plot the
      result of our simulation, in which
      reionization occurs at $z_{\rm re}=11.5$, a redshift at which the HOF
      is expected to become largely independent of the exact value of $z_{\rm re}$. {\em Bottom panel}: the present-day halo mass at which the halo occupation fraction, HOF=$0.5$, as a function of $z_{\rm re}$. If galaxies form though atomic hydrogen cooling prior to reionization, then the HOF becomes independent of the value of $z_{\rm re}$ for both, $z_{\rm re} > 10$ and $z_{\rm re} < 2$.}
    \label{Fig:zr_HOF}
\end{figure}

Fig.~\ref{Fig:zr_HOF} shows the result of this calculation. The HOF is clearly sensitive to the
assumed value of $z_{\rm re}$ unless reionization occurred either early or very late. For
example, the bottom panel of Fig.~\ref{Fig:zr_HOF} shows that for $z_{\rm re} > 10$ (thick line in the top panel), the halo mass at which only half the
dark matter haloes host a luminous galaxy is
$M_{200,50} \sim 10^{9.4} M_{\odot} \ h^{-1}$. On the other hand, if
the universe is reionized at $z=4$, the value of $M_{200,50}$ is a
factor of 4 lower. In the limiting case of a universe that experiences very late reionization (or no reionization), the assumption that galaxies form through atomic hydrogen cooling yields
$M_{200,50} \sim 10^{8.6} \ M_{\odot} h^{-1}$. For comparison, the left red
dashed line in the top panel shows the HOF measured in a zoom-in simulation of the
Local Group carried out by~\cite{Sawala2016} that does not include
reionization. Reassuringly, our model reproduces their simulation
results very well. We also plot the result of our own simulation, in which cosmic reionization occurs at
$z_{\rm re} = 11.5$, i.e., early enough that the measured HOF is
independent of the exact value of $z_{\rm re}$.

We conclude that if galaxy formation is largely determined by atomic hydrogen cooling prior to cosmic reionization, then: 
\begin{itemize}
    \item{the $z = 0$ HOF is insensitive to the redshift of reionization
        provided $z_{\rm re} > 10$ and $z_{\rm re} < 2$.} 
    \item{the $z=0$ HOF depends strongly on the redshift of reionization if $2 < z_{\rm re} < 10$.} 
\end{itemize}

The model of Section~\ref{Sec:Model} thus provides
a simple framework to account for both the qualitative and
quantitative effect of $z_{\rm re}$ on the HOF. We now turn our attention
to the impact of varying the critical mass for galaxy formation prior
to reionization.
 
\subsection{Pre-reionization galaxy formation and the Halo
  Occupation Fraction}
\label{Sec:impact_of_early_gf}

We now investigate the sensitivity of the HOF to assumptions about
how galaxy formation proceeds prior to reionization. For the purposes of this discussion we fix the redshift of reionization
to $z_{\rm re} = 11.5$.

\begin{figure}
    \centering
    \includegraphics{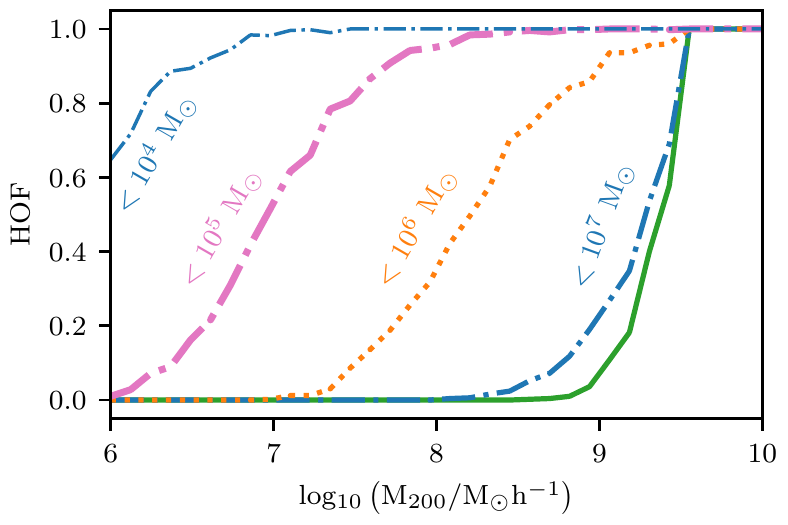}
    \caption[]{{\it Halo Occupation Fraction} (HOF) at the present day
      as a function of halo mass predicted by our model
      (Section~\ref{Sec:Model}), assuming a redshift of
      reionization, $z_{\rm re} = 11.5$. The green solid line shows the
      limiting case of assuming that {\it all} dark matter haloes reach $z_{\rm re}$ without hosting a luminous galaxy; the other curves assume that {\it all} haloes that exceeded the quoted mass prior to reionization host a galaxy by $z_{\rm re}$. The $z=0$ HOF depends strongly on
      assumptions about galaxy formation prior to reionization, but
      only if galaxy formation proceeds in haloes well below the
      atomic hydrogen cooling limit at $z_{\rm re}$. For $z_{\rm re} \sim 11.5$,
      this corresponds to $M_{200} \sim 10^{7} h^{-1} M_{\odot}$.}
    \label{Fig:who_is_dark_MAH}
\end{figure}

Fig.~\ref{Fig:who_is_dark_MAH} shows the $z= 0$ HOF obtained from
our model, using our set of EPS assembly histories for
different assumptions. 
Specifically, we consider the limiting case where {\it all} dark
matter haloes reach $z_{\rm re}$ without hosting a luminous galaxy (green
solid curve), as well as the cases when galaxy formation before $z_{\rm re}$
occurs in halos of mass $M_{200} > 10^{7} M_{\odot} h^{-1}$
(blue dot-dashed line), $10^{6} M_{\odot} h^{-1}$ (orange dotted
line), $10^{5} M_{\odot} h^{-1}$ (pink dot-dashed line), and
$10^{4} M_{\odot} h^{-1}$ (blue dot-dashed line). Not surprisingly, these
different assumptions about pre-reionization galaxy formation have a dramatic impact on the HOF.

The critical mass for atomic hydrogen cooling to become efficient at
$z=11.5$ is $M_{H}^{11.5} \sim 3\times 10^{7} M_{\odot}
h^{-1}$. For this value of $z_{\rm re}$, the present-day HOF varies little
if we assume that {\it all} dark matter haloes reach $z_{\rm re}$ without a
luminous galaxy in them or only those with $M_{200} \sim
M_{H}^{z_{\rm re}}$. However, if we allow less massive haloes to be populated
with luminous galaxies prior to reionization - such as haloes of mass
$M_{200} = 10^{6} M_{\odot} h^{-1}$ at $z_{\rm re}$ - then the HOF becomes
very sensitive to our assumptions. 

One caveat of the exercise carried out in Fig.~\ref{Fig:who_is_dark_MAH} is that the halo occupation fraction prior to reionization may not be a simple step function. Although the formation of the first galaxies depends primarily on halo mass, it may also depend on secondary parameters possibly related to the local environment~\cite[e.g.][and references therein]{Couchman1986, Yoshida2003, Yoshida2006}. Recent numerical simulations by~\cite{Wise2014} show, however, that although the halo occupation fraction is not a step function prior to reionization, the transition between luminous and starless haloes occurs over a very narrow range of halo mass at $z=15$ (roughly a factor of 3 in halo mass below the atomic hydrogen cooling limit).

We will next show that if the redshift of reionization is lower than $z_{re} = 10$, the HOF depends quite strongly on the assumptions about galaxy
formation before reionization whereas if the redshift is higher, the converse is true.

\subsection{Combined impact of $z_{\rm re}$ and pre-reionization galaxy formation}

\begin{figure}
    \centering
    \includegraphics[width=\columnwidth]{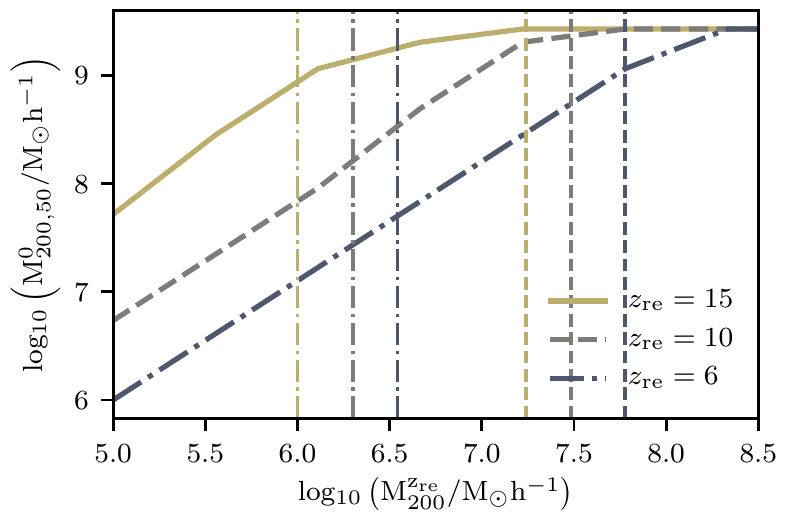}
    
    \caption[]{Halo mass at $z= 0$ at which $50 \%$ of
      haloes host a galaxy, as a function of the halo mass at $z_{\rm re}$
      above which {\it all} halos are assumed to have become luminous
      prior to $z_{\rm re}$. Different colours correspond to different
      values of the redshift of reionization, $z_{\rm re}$, as indicated in the legend. As an example, if {\it all} haloes more massive than
      $10^{6}~M_{\odot}$ already host a luminous galaxy by
      $z_{\rm re} = 10$, then $50 \%$ of the halos with halo mass,
      $M_{200} \sim 6 \times 10^7 \ M_{\odot}$ at $z=0$ will also
      host a luminous galaxy. The vertical dashed lines show the
      critical atomic cooling mass at $z_{\rm re}$ (i.e., lines of constant
      temperature, $T_{200} = 8000 K$). Vertical dot-dashed
      lines correspond to a constant virial temperature,
      $T_{200} = 1000 K$.}
    \label{Fig:impact_reionization}
\end{figure}

We now explore how the $z=0$ HOF depends on variations of both the
assumed redshift of reionization, $z_{\rm re}$, and the
pre-reionization critical mass for galaxy formation,
$M_{c}^{z_{\rm re}}$. For each value of $z_{\rm re}$, we compute the
HOF by comparing the individual EPS assembly histories to the critical
mass derived from our model (Section~\ref{Sec:Model}), assuming
that haloes more massive than $M_{c}^{z_{\rm re}}$ at $z_{\rm re}$ are
already luminous at $z_{\rm re}$. We express the outcome of this
calculation as the halo mass at $z=0$ at which only half of the dark
matter haloes are luminous, $M_{200,50}^{0}$, as a function of both, $z_{\rm re}$ and
$M_{c}^{z_{\rm re}}$. This is shown in
Fig.~\ref{Fig:impact_reionization} for three different assumed values of $z_{\rm re}$.

Fig.~\ref{Fig:impact_reionization} shows that both $z_{\rm re}$ and
$M_{c}^{z_{\rm re}}$ affect the characteristic mass scale at which $50 \%$ of
the haloes are luminous at $z=0$. As discussed in the previous
section, if reionization occurred early, the
characteristic halo mass at $z=0$ at which $50 \%$ of the haloes
remain dark is roughly independent of assumptions about early galaxy
formation, as long as galaxies form in haloes more massive than
$10^{6} M_{\odot}$ prior to $z_{\rm re}=15$. However, if reionization occurs
much later, then the critical halo mass for galaxy formation before
reionization has a much greater impact on the $z=0$ HOF.

Detailed cosmological simulations of the formation of the first
galaxies - which include cooling processes, such as molecular hydrogen
cooling, ignored by our simulation - seem to support the idea that the
process of galaxy formation is largely triggered only in haloes in
which atomic hydrogen cooling becomes efficient, i.e., those more
massive than $2 \times 10^{7} \ M_{\odot} h^{-1}$ at $z\sim 15$~\citep[see
e.g.,][and references therein]{Xu2013, Xu2016}. For such high redshift
of reionization, our results are already robust even if we allow
haloes 10 times smaller than this to host a galaxy by that time. 

If the redshift of reionization is much lower, $z=6$ say, then the
critical halo mass at which $50\%$ of the dark matter haloes host a
galaxy by today becomes independent of the exact assumptions about
early galaxy formation, provided galaxy formation occurs in haloes
more massive than $\sim 10^{8} M_{\odot} h^{-1}$ before
reionization. This mass is not far from the atomic hydrogen cooling
limit at $z_{\rm re}$ (vertical green dashed line), but allowing
low-mass haloes with virial temperature $T_{200} \sim 1000 K$
(vertical dot-dashed lines) to host a galaxy at $z_{\rm re}$, has a
large impact on the $z=0$ HOF.

We conclude that the lower the redshift of reionization, the more dependent the HOF is on assumptions about pre-reionization galaxy formation. Neglecting cooling processes that are important below a virial temperature of $10^{4} K$, as we have done, might affect the expected
$z=0$ HOF if reionization occurred as late as $z\leq 6$.

\section{DISCUSSION}
\label{Sec:Discussion}

The agreement between the model of Section~\ref{Sec:Model} and our
high-resolution cosmological hydrodynamic simulation suggests that the
onset of galaxy formation in low-mass haloes, as well as the detailed structure of RELHICs, can be
understood in detail in the context of the $\Lambda$CDM cosmology. The key
inputs are: {\it (i)} the redshift at which hydrogen was reionized,
{\it (ii}) the redshift-dependent temperature of the intergalactic
medium after reionization (which is largely set by the external UVB
radiation field), and {\it (iii)} the minimum halo mass in which
galaxies can form prior to reionization. Some of the predictions of
our model are relatively insensitive to the details of these inputs.

If reionization occurred in the redshift range,
$6 < z_{\rm re} < 10$, as indicated by recent Planck
results~\citep{Planck2018}, then the minimum halo mass today that can
host a luminous galaxy cannot be much smaller than
$M_{200} \sim 10^8 M_{\odot} h^{-1}$, unless galaxy formation was able
to take place in haloes of virial temperature
$T_{200} < 2 \times 10^{3}K$ prior to reionization (or
$M_{200} \lesssim 3 \times 10^{6} M_{\odot} h^{-1}$ at
$z_{\rm re} = 10$). Recent hydrodynamical simulations indicate that
this is unlikely \citep{Xu2016}.  Thus, the
$M_{200} \sim 10^8 M_{\odot} h^{-1}$ mass scale should be a general
outcome of simulations of galaxy formation, regardless of the details
of the physical modelling. 

\cite{Shen2014} studied the formation of dwarf galaxies in seven halos
in the mass range
$3 \times 10^{8} \lesssim M_{200} \ M_{\odot} h^{-1} \lesssim 2.4
\times 10^{10}$, in high-resolution zoom simulations carried out with
the {\tt GASOLINE} code. They assumed the~\cite{Haardt2012} UVB
external radiation field and a redshift of reionization,
$z_{\rm re} = 6$. They found that three of their haloes, of mass less
than $M_{200} \lesssim 4.5 \times 10^{8} M_{\odot} h^{-1}$, remained
{\it starless} at $z=0$. They also found that the gas density in these
``dark'' haloes had never exceeded
$n_{\rm H} \sim 10^{-2} \rm \ cm^{-3}$, a value well below the imposed
density threshold for star formation in their simulations
($n_{\rm H} = 100 \rm \ cm^{-3}$). These results are in excellent
agreement with ours and are readily understood in terms of our model
of Section~\ref{Sec:Model}. In this mass range, our model predicts a
maximum central gas density almost two orders of magnitude lower than
that reached in haloes that are close to the critical mass (see bottom
panel of Fig.~\ref{Fig:structural}).

The same minimum halo mass for galaxy formation also emerges from the
high-resolution zoom cosmological simulations of dwarf galaxies
carried out by~\cite{Munshi2019}, also with the {\tt GASOLINE} code,
but with different assumptions for star formation. They found that
both simulations that include, and simulations that ignore molecular
hydrogen cooling make galaxies only in {\it field} haloes more massive
than $M_{200} \gtrsim 3 \times 10^{8} M_{\odot} h^{-1}$ at $z=0$, as
predicted by our model. The authors find that neglecting molecular
hydrogen cooling produces twice as many luminous galaxies as when this
cooling process is included. This counterintuitive result, which is at
odds with the expectations of our model, may simply reflect the
particular treatment of star formation in the latter simulations,
which is closely tied to the fraction of molecular hydrogen. There is
observational~\citep[e.g.,][]{Michalowsky2015} and
theoretical~\citep[e.g.][]{Hu2016} evidence suggesting that the
presence of molecules is a sufficient, but not a necessary condition
for star formation, which is mostly regulated by atomic hydrogen.  For
example, the high-resolution cosmological hydrodynamical simulations
of \cite{Xu2013} that include radiative transfer, non-equilibrium
chemistry and cooling by molecular hydrogen are consistent with the
onset of galaxy formation being regulated mainly by atomic, rather
than molecular hydrogen cooling prior to reionization.

Thus, our results, together with those from other authors, provide
strong evidence for the existence of a minimum halo mass below which
galaxy formation cannot proceed. The descendents of these barren
haloes have mass, $M_{200} \lesssim 10^{8} M_{\odot} h^{-1}$, at
$z=0$. Finding a luminous galaxy in a dark matter halo close to this
mass scale should be rare, but not impossible.  For example, our
simulation indicates that there should be
$\sim 3.4 \times 10^{-2} \rm / Mpc^{3}$ luminous galaxies inhabiting
haloes of mass $M_{200} \lesssim 10^{9} M_{\odot} h^{-1}$.  Claims
challenging the $\Lambda$CDM model on the grounds that there are
nearby {\it field} dwarfs surrounded by low-mass haloes of mass
$M_{200} \lesssim 10^{10} \ M_{\odot}$~\citep{Ferrero2012} should
therefore be treated with caution.

Finally, our results also indicate that {\it field} dark matter haloes
above the mass scale,
$M_{200} \gtrsim 5 \times 10^{9} M_{\odot} h^{-1}$, should always host
a luminous galaxy at $z=0$. As far as we are aware, no simulation of
galaxy formation has ever found a dark matter halo with mass above
this mass scale that remains devoid of stars at $z=0$~\cite[e.g.][and
references therein]{Benitez-Llambay2015, Sawala2016, Fitts2017,
  Munshi2019}.

\section{SUMMARY AND CONCLUSIONS}
\label{Sec:Conclusions}

We have developed a simple theoretical framework, in the context of
the $\Lambda$CDM cosmological model, to follow the formation of the
first galaxies before and after the reionization of hydrogen at early
times. In particular, we have introduced a simple analytical model to
calculate the properties of the gas component of haloes after
reionization. Within this framework we have addressed two specific
problems: the mass of the smallest haloes that can sustain galaxy
formation, and the properties of RELHICs -{\it starless} gaseous haloes whose gas was unable
to collapse by the time the universe reionized-. We introduced the
concept of the {\it Halo Occupation Fraction} (HOF), which we have
defined as the fraction of dark matter haloes that host a luminous
galaxy today as a function of halo mass.

In our model galaxy formation can only take place in haloes whose mass
exceeds a redshift-dependent critical value, $M_{\rm cr}^z$. Before
reionization, $M_{\rm cr}^z$ is the mass at which atomic hydrogen
cooling becomes effective, $\sim 4.8 \times 10^7 \ M_{\odot}$ at $z =7$, for
example. After reionization, $M_{\rm cr}^z$ is the mass above which
gas, which was previously kept in equilibrium in a halo by the
pressure of the UVB radiation, can no longer remain in hydrostatic
equilibrium and therefore collapses to make a galaxy.  The HOF is then
the result of the interplay between the evolution of this critical
mass and the assembly history of haloes. It is completely determined
by three factors: the minimum halo mass for galaxy formation before
reionization; the redshift of reionization, $z_{\rm re}$; and the
intensity of the (evolving) external photoheating rate.

The properties of haloes whose gaseous component has not yet collapsed
to make a luminous galaxy by $z_{\rm re}$ are determined by the
effects of reionization. After reionization, the temperature of the
gas in the high-density intergalactic medium is set by the balance
between radiative cooling and photoheating and, in low density gas, by
photoheating alone.  We developed a simple analytical model to study
how the properties of the ionized intergalactic medium affects the
properties of gaseous haloes as a function of time.

We validated our model by comparing with results of a cosmological hydrodynamics
simulation that models the main astrophysical processes involved in galaxy formation.
The main conclusions arising from our model may be summarised as follows.

\begin{itemize}

\item{If, prior to reionization, galaxies form predominantly from gas
    that condenses through atomic hydrogen cooling, then the HOF at
    $z=0$ is almost insensitive to $z_{\rm re}$ provided
    $z_{\rm re} > 10$.  We tested two limiting cases, $z_{\rm re} > 10$
    and $z_{\rm re} << 10$, with our high-resolution cosmological
    hydrodynamical simulation and found that the HOF agrees remarkably
    well with the predictions of our model (see
    Fig.~\ref{Fig:zr_HOF}).}

\item{If the redshift of reionization is $z_{\rm re}>>10$, then the
    $z=0$ HOF becomes insensitive to the details of how gas cools within
    haloes before reionization. For example, if $z_{\rm re} = 15$, and
    galaxies were able to form before this redshift in haloes more
    massive than $10^{6} M_{\odot} h^{-1}$ (in which molecular
    hydrogen cooling could be relevant) there would be very little change in the $z=0$ HOF.}

\item{If the redshift of reionization is $z_{\rm re} < 10$, the $z=0$ HOF
    depends sensitively on the minimum halo mass that can host a
    luminous galaxy before reionization. For example, if gas cooled
    mainly by molecular (rather than atomic) hydrogen processes in
    haloes of virial temperature $T_{200} \sim 10^3 K$, then the mass
    scale at which 50\% of the haloes are luminous at $z=0$ can vary
    by more than a factor of 10 compared to the atomic hydrogen 
    cooling case (see, e.g., Fig.~\ref{Fig:impact_reionization}).}

\item{The potential well of dark matter haloes less massive than
    $M_{200} \lesssim 10^{6} M_{\odot} h^{-1}$ at $z=0$ is not deep
    enough to overcome the pressure force produced by the
    external~\cite{Haardt2001} UVB radiation field. The gas completely evaporates from these haloes which are therefore gas free at $z=0$. They will
    also be star-free unless galaxy formation occurred in haloes of mass $M_{200} \sim 10^{4} M_{\odot} h^{-1}$ prior
    $z_{\rm re} > 10$, which is unlikely.}

\item{The gas content of dark matter haloes in the mass range
    $10^6 \lesssim M_{200} / M_{\odot} h^{-1} \lesssim 5\times 10^{9}$
    at $z=0$ is drastically reduced (but not completely removed) by the presence of the UVB radiation field. A small fraction of the
    remaining gas is prevented from collapsing by the external
    radiation field, with which it is in thermal equilibrium, and is
    confined, in hydrostatic equilibrium, within the potential well of
    the halo, without forming stars. These are the objects we
    previously identified in cosmological hydrodynamics simulations
    and called RELHICs (Reionization limited HI clouds; 
    \citealt{Benitez-Llambay2017}).  The gas density profiles of RELHICs
    are flat at large distances from the halo centre and have a
    well-defined maximum value at the centre; this latter property is
    a direct consequence of the gas being in equilibrium in the cuspy
    dark matter distribution of $\Lambda$CDM haloes.}

\item{The amount of gas required to overcome the gravitational
    acceleration of dark matter haloes more massive than
    $M_{200} \sim 5 \times 10^{9} M_{\odot} h^{-1}$ exceeds the
    universal baryon fraction ($\Omega_{b}/\Omega_{m}$) by orders of
    magnitude. This gas cannot remain in hydrostatic equilibrium and
    will therefore collapse to make a luminous galaxy.}

\end{itemize}

We tested the results of our analytical model by comparing with
cosmological hydrodynamics simulations. We find that our predicted HOF
is reproduced remarkably well in the simulations. Similarly, the gas
mass and internal structure of RELHICs predicted by our model are
reproduced very well in the simulations. 

Our results apply strictly to {\it field} (or {\em central}) dark
matter haloes, not to {\it satellites} of more massive systems. Unlike
field haloes whose mass can only grow with time, satellites can
lose mass by tidal stripping. Note, however, that our model applies to
satellites before they fell into their host halo.

Our simulation indicates that there should be fewer than $30$ {\it
  field} luminous galaxies inhabiting dark matter haloes of mass
$M_{200} \lesssim 10^{9} M_{\odot} h^{-1}$ in the Local
Volume\footnote{We take the Local Volume to be sphere of $10$ Mpc
  centred on the Milky Way, and assume that our simulation is
  representative of the Local Volume.}. Finding many more nearby {\it
  field} dwarfs in such low-mass haloes would require either a
revision of our current understanding of how (and when) reionization
occurred or of how galaxy formation proceeded before
reionization. Interestingly, \cite{Jethwa2018} and~\cite{Graus2019}
have recently claimed, based on simulations, that such a population of
galaxies might be required to explain the sheer number and radial
distribution of ultrafaint galaxies observed around the Milky Way but
\cite{Bose2019} argue that these results are affected by numerical
resolution and that, when these effects are taken into account, the
data are in good agreement with the observed number of ultrafaint
satellites. In any case, our model provides a theoretical framework to address these issues and
interpret that data without the need for cosmological simulations and
thus, independent of computational cost or resolution limitations.

In summary, if the onset of galaxy formation is largely determined by
atomic hydrogen cooling in small haloes before reionization, then we
expect a cutoff in the galaxy mass function at a halo mass,
$M_{200} \sim 10^{8} M_{\odot} h^{-1}$, at $z=0$. Finding an isolated
luminous galaxy in a dark matter halo below this mass scale would be a
remarkable discovery that would challenge not only the current
cosmological model, but also current understanding of galaxy
formation.

\section{Acknowledgements}
We are particularly grateful to Tom Theuns, John Regan and John Wise for providing detailed comments on an earlier version of our paper. These helped to improve our presentation. We also acknowledge the anonymous referee for their report which helped to improve the current manuscript. We acknowledge support from the European Research Council
through ERC Advanced Investigator grant, DMIDAS [GA 786910] to CSF. This work was also supported by STFC Consolidated Grants for Astronomy at Durham ST/P000541/1 and ST/T000244/1. This work used the DiRAC Data Centric system at Durham University, operated by the Institute for Computational Cosmology on behalf of the STFC DiRAC HPC Facility (www.dirac.ac.uk). This equipment was funded by BIS National E-infrastructure capital grants ST/P002293/1, ST/R002371/1 and ST/S002502/1, Durham University and STFC operations grant ST/R000832/1. DiRAC is part of the National e-Infrastructure. The simulation used in this work was performed using DiRAC's director discretionary time. We are grateful to Prof. Mark Wilkinson for awarding this time. 

\bsp	
\label{lastpage}

\section*{Data availability}
The data underlying this article will be shared on reasonable request to the corresponding author.
\bibliographystyle{mnras}
\bibliography{bibliography} 

\appendix
\section{Effective density-temperature relation of star-free gaseous haloes}
\label{App:eos}

Table~\ref{Tab:eos} summarises the density-temperature relation used
in this work which is derived from the~\cite{Wiersma2009} cooling
tables.

\begin{table*}
    \centering
\begin{tabular}{crrrrrrrrrrrrrrrrr}
\hline
$\rm log_{10} \left ( n_{H} / cm^{-3} \right )$ & -8.0 & -7.5 & -7.0 & -6.5 & -6.0 & -5.5 & -5.0 & -4.5 & -4.0 & -3.5 & -3.0 & -2.5 & -2.0 & -1.5 & -1.0 & -0.5 &  0.0 \\
\hline
\hline
$z=0.0$ & 2.89 & 3.16 & 3.45 & 3.73 & 4.02 & 4.28 & 4.49 & 4.54 & 4.43 & 4.30 & 4.20 & 4.13 & 4.08 & 4.03 & 4.00 & 3.97 & 3.94 \\
$z=1.1$ & 2.83 & 3.00 & 3.23 & 3.49 & 3.76 & 4.03 & 4.29 & 4.51 & 4.63 & 4.53 & 4.40 & 4.29 & 4.20 & 4.14 & 4.09 & 4.04 & 4.00 \\
$z=2.0$ & 2.79 & 2.91 & 3.10 & 3.33 & 3.59 & 3.86 & 4.13 & 4.37 & 4.59 & 4.57 & 4.47 & 4.36 & 4.26 & 4.18 & 4.12 & 4.07 & 4.03 \\
$z=3.0$ & 2.81 & 2.89 & 3.03 & 3.24 & 3.49 & 3.75 & 4.02 & 4.27 & 4.51 & 4.54 & 4.44 & 4.33 & 4.23 & 4.17 & 4.11 & 4.06 & 4.02 \\
$z=4.1$ & 2.78 & 2.84 & 2.96 & 3.15 & 3.39 & 3.65 & 3.93 & 4.20 & 4.43 & 4.49 & 4.39 & 4.29 & 4.21 & 4.15 & 4.09 & 4.05 & 4.01 \\
$z=5.2$ & 2.70 & 2.76 & 2.88 & 3.07 & 3.31 & 3.57 & 3.84 & 4.13 & 4.35 & 4.43 & 4.35 & 4.26 & 4.18 & 4.13 & 4.08 & 4.03 & 4.00 \\
$z=6.1$ & 2.60 & 2.69 & 2.84 & 3.03 & 3.27 & 3.53 & 3.79 & 4.05 & 4.27 & 4.35 & 4.31 & 4.23 & 4.16 & 4.11 & 4.06 & 4.02 & 3.99 \\
$z=7.2$ & 2.44 & 2.54 & 2.72 & 2.93 & 3.19 & 3.45 & 3.71 & 3.95 & 4.15 & 4.29 & 4.27 & 4.21 & 4.15 & 4.10 & 4.05 & 4.01 & 3.98 \\
$z=8.0$ & 2.28 & 2.42 & 2.60 & 2.85 & 3.11 & 3.37 & 3.63 & 3.85 & 4.03 & 4.21 & 4.25 & 4.19 & 4.14 & 4.09 & 4.04 & 4.01 & 3.98 \\
$z=9.0$ & 2.04 & 2.24 & 2.48 & 2.73 & 3.03 & 3.27 & 3.51 & 3.71 & 3.91 & 4.11 & 4.22 & 4.18 & 4.12 & 4.08 & 4.03 & 4.00 & 3.97 \\
\hline
\end{tabular}
    \caption[]{{\it Temperature-density} relation at different
      redshifts derived from the~\cite{Wiersma2009} cooling tables,
      which assume an optically thin gas in ionization equilibrium,
      illuminated by the~\cite{Haardt2001} UVB radiation
      field. For low-density gas the relation is such that the photoheating timescale equals the Hubble time.
      For high-density gas, the temperature is such that the
      cooling rate equals the photoheating rate. Entries in the table are $\rm log_{10}(T/K)$.} 
    \label{Tab:eos}
\end{table*}

\section{Is galaxy formation triggered when gaseous haloes exceed the critical mass?}
\label{App:when_galaxies_form}

\begin{figure}
    \centering
    \includegraphics[width=\columnwidth]{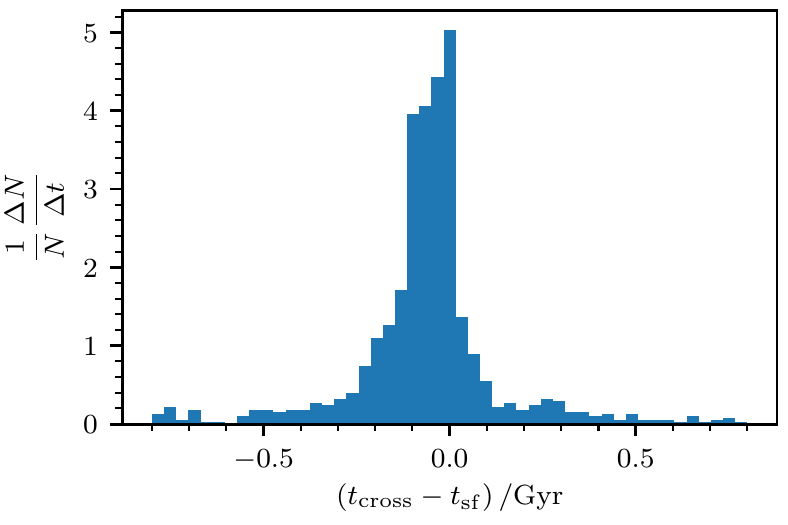}
    \caption[]{Distribution of $t_{\rm cross}-t_{sf}$, i.e., the difference of the time at which a simulated luminous dark matter halo first crossed the critical mass for galaxy formation to take place ($t_{\rm cross}$), and the time at which the first star was formed in the halo ($t_{\rm sf}$). As expected, most haloes form their stars soon after they become more massive than the critical mass. There is, however, a large population of haloes for which star formation occurs slightly after they cross the critical mass. This delay, which is a negligible fraction of the Hubble time at $z=0$, is likely due to the time that it takes for the gas to collapse after it becomes self-gravitating.}
    \label{Fig:app_when_galaxies_form}
\end{figure}

Galaxy formation in our simulation typically takes place soon after the RELHICs exceed the critical mass for gas to lose pressure support. We show this in Fig.~\ref{Fig:app_when_galaxies_form}, which displays the distribution of the difference, $t_{\rm cross}-t_{\rm sf}$, i.e., the difference between the cosmic time at which the simulated dark matter haloes that host a luminous galaxy at $z=0$ first crossed the critical mass for galaxy formation to take place ($t_{\rm cross}$), and the time at which the first star was formed in the halo ($t_{\rm sf})$. The  difference is close to 0, indicating that the process of galaxy formation is indeed linked to the event of exceeding the critical mass. The small bias towards negative values (i.e., $t_{\rm sf} \gtrsim t_{\rm cross}$) is likely due to the time that it takes for the gas to undergo gravitational collapse before turning into stars, which is typically a few hundred $\rm Myr$.

\section{EPS mass assembly histories}
\label{App:MAHs}
\begin{figure}
    \centering
    \includegraphics{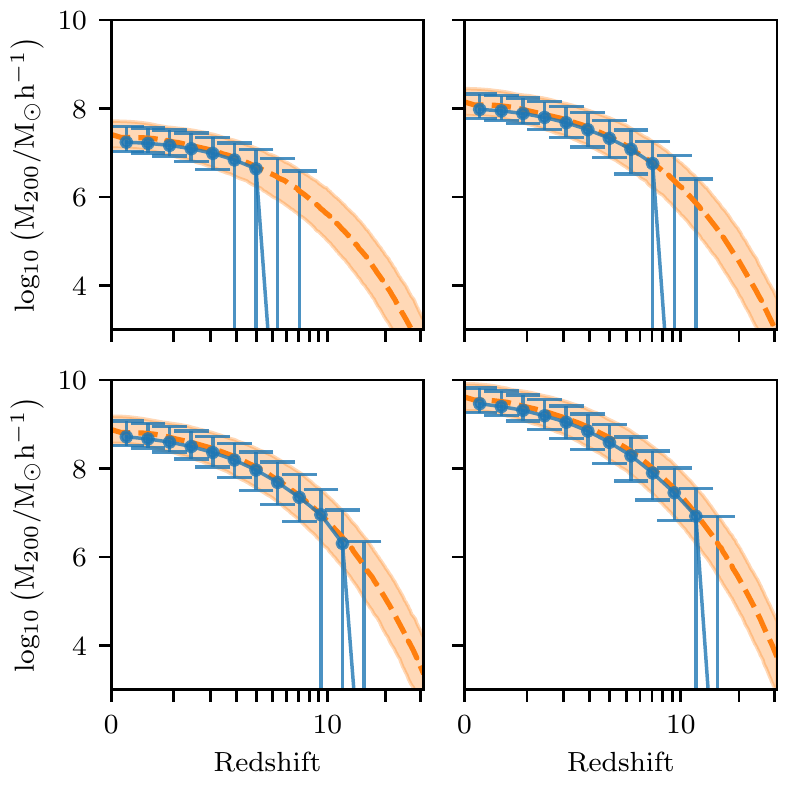}
    \caption[]{Comparison of synthetic assembly histories constructed
      using the Extended-Press-Schechter (EPS) formalism (orange
      dashed line) and assembly histories measured directly in our
      simulation (blue dots), for four different present-day mass
      bins. The shaded region shows the (10-90)th percentiles
      around the median for the EPS assembly histories; the error bars show the same quantities for the simulated dark matter
      haloes.}
    \label{Fig:synthetic_MAHs}
\end{figure}

Fig.~\ref{Fig:synthetic_MAHs} shows the comparison between the EPS
assembly histories (orange dashed lines) and those constructed
directly from the simulation using {\tt HBT+} (blue dots) for four bins of present-day halo mass. The shaded regions
indicate the (10-90)th percentiles of the EPS assembly histories and the
error bars the corresponding quantity for the simulated haloes. The
EPS assembly histories agree well with those measured in the
simulation down to the resolution limit of the simulation
($M_{200} \sim 10^{6} M_{\odot} h^{-1}$). Furthermore, as shown in
Fig.~\ref{Fig:simulated_HOF}, the HOF derived using the EPS assembly histories (green dot-dashed line) agrees remarkably well with the HOF measured in the simulation.

\end{document}